\documentclass{jfm}
\usepackage{graphicx}
\usepackage{float}
\usepackage{xcolor,soul}
\usepackage{epstopdf, epsfig}
\usepackage{amsmath}
\usepackage{multirow}
\usepackage{placeins}
\usepackage{caption}
\usepackage{changepage}
\usepackage{booktabs}
\usepackage{bm}
\usepackage{makecell}
\usepackage[numbers]{natbib}

\shorttitle{Experimental SDRL-based control of TS waves}

\shortauthor{Mohammadikalakoo et al.}

\title{Experimental Demonstration of SDRL Controller for TS Wave Suppression with DBD Actuator}

\author{B. Mohammadikalakoo\aff{1}
  \corresp{\email{b.mohammadikalakoo@tudelft.nl}},
  S. G. Villasol\aff{2},
  G. Salomone\aff{3},
  M. Kotsonis\aff{1}
 \and  N. A. K.  Doan\aff{4,1}}

\affiliation{
\aff{1} Department of Flow Physics and Technology, Faculty of Aerospace Engineering, Delft University of Technology, Kluyverweg~1, 2629~HS~Delft, The Netherlands
\\[3pt]
\aff{2} Department of Aerospace Engineering, Universidad Carlos~III~de~Madrid, Avenida de la Universidad~30, 28911~Leganés, Spain
\\[3pt]
\aff{3} Dipartimento di Ingegneria Industriale, Università degli Studi di Napoli “Federico~II”, Piazzale Vincenzo Tecchio~80, 80125~Napoli, Italy
\aff{4} Department of Aeronautics, Imperial College London, SW7 2AZ London, United Kingdom
}

\begin{document}
\maketitle

\begin{abstract}

An experimental wind-tunnel implementation of a model-free single-step deep reinforcement learning (SDRL) controller is presented for TS wave suppression in a flat plate boundary layer. The controller is deployed in a feedforward layout. The arrangement comprises an upstream reference microphone, a downstream error microphone, and a DBD plasma actuator located between them. The controller updates its policy online from the measured error signal and, in real time, adjusts the coefficients of a finite-impulse-response (FIR) filter that maps the reference signal to the actuation command. TS waves are artificially introduced by a second, upstream-located DBD “trigger” actuator identical in specification to the control actuator. The trigger actuator is driven with single-frequency, multi-frequency, or broadband white-noise inputs depending on the control cases. Experiments were carried out in an anechoic wind tunnel facility using flush-mounted pressure microphones for sensing and controller feedback, together with two-component planar particle image velocimetry~(PIV) for flow-field verification. The controller performance is assessed via second-order statistics of the error signal and the spectral attenuation of the TS wave content. Across all tested scenarios, the SDRL-based controller consistently reduces the downstream disturbance level and exhibits robustness to moderate variations in freestream velocity and in the incoming TS wave disturbance spectrum. These results provide an experimental step toward adaptable, data-driven TS wave suppression with compact sensing and actuation, supporting practical strategies for boundary layer transition delay.
\end{abstract}


\newpage
\section{Introduction}
Reduction of aerodynamic drag remains a central challenge in aerospace engineering because it directly influences fuel consumption, emissions, and overall efficiency. In typical transport configurations, skin-friction drag can account for up to half of the total aerodynamic resistance~\citep{Walther2001OptimalLayer}. Because skin-friction drag arises from the wall shear stress, which increases sharply after transition to turbulence, delaying the laminar-turbulent transition offers an effective route toward improving aerodynamic performance by skin-friction drag reduction. The transition in two-dimensional low-disturbance boundary layers is primarily triggered by the amplification and breakdown of Tollmien-Schlichting (TS) waves~\citep{Schlichting2017}.

Efforts to suppress these convective instabilities have historically relied on two main categories of flow-control strategies: passive and active~\citep{Gad3}. Passive methods, such as distributed roughness elements~\citep{Fransson2004,fransson2006}, compliant surfaces~\citep{carpenter85,davies1997}, or resonance-based devices like phononic crystals and Helmholtz resonators~\citep{hussein2015,michelis2023,michelis2023_2}, modify the boundary layer receptivity or stability characteristics without energy input. Although effective in controlled environments, these approaches lack adaptability to varying flow conditions. 

Active flow control (AFC), by contrast, introduces energy into the flow via steady or unsteady actuation. Predetermined active strategies, such as uniform suction and blowing~\citep{saric1983effect}, surface heating~\citep{Liepmann1946,Nosenchuck1982}, Lorentz forcing~\citep{albrecht2006}, or dielectric-barrier discharge (DBD) plasma actuation~\citep{roth98,duchmann2013,dorr2018}, alter the base flow in open loop. While these methods can delay transition, their open-loop nature prevents compensation for environmental or operational variations. To achieve robustness, reactive (closed-loop) schemes have been developed, where real-time sensor feedback drives actuators to cancel incoming TS waves through wave superposition~\citep{Thomas,superposition3,fabbiane2017}. These reactive methods proved more energy-efficient and adaptive to flow changes, motivating subsequent developments in feedback control theory.

Modern AFC research distinguishes between model-based and model-free reactive control. Model-based approaches (e.g., linear–quadratic–Gaussian~(LQG), model predictive control~(MPC), and inverse feedforward control~(IFFC)) rely on reduced-order representations of the Navier–Stokes equations to compute optimal actuation laws~\citep{Bagheri2009-input,belson2013,Brito2021,tol2019experimental}, offering precision at the expense of robustness to model uncertainty. Model-free strategies, instead, adapt directly from sensor data, most prominently via the filtered-$x$ least mean squares (FXLMS) algorithm~\citep{Ladd1988,baumann1996,fabbiane2014,Marios2015}, which continuously adjusts control parameters without requiring explicit plant identification. These adaptive feedforward systems have been demonstrated experimentally with plasma actuators for both monochromatic and broadband TS disturbances~\citep{modelfree2,modelfree3,mohammadikalakoo_paper}.

While classical model-free controllers offer practical robustness, their adaptation is typically confined to linear input–output models with linear or quadratic update laws, which can limit the ability to capture complex temporal dependencies and result in slow convergence. This motivates learning-based methods that capture nonlinear, long-horizon sensor–actuator relationships. Early neural-network-based controllers introduced nonlinear input–output mappings for flow control~\citep{fan1993active,fan1995transition}. Among a flurry of artificial intelligence techniques used recently in fluid dynamics applications, deep reinforcement learning (DRL) has shown strong potential across both laminar~\citep{rabault2019artificial, rabault2019accelerating, Paris2021, li2022, chen2023, wang2023, mohammadikalakoo_paper, mohammadikalakoo_canada, Hu2024} and turbulent~\citep{ren2021, guastoni2023, Font2025} regimes by autonomously exploring and refining control policies from feedback. Nevertheless, full DRL pipelines can be impractical in fluid mechanics: they often require high-dimensional observations, large datasets, substantial computing resources, and prolonged interaction with expensive solvers, together with nontrivial network design and tuning. To alleviate these burdens, efficient variants have been proposed, such as DRL coupled with low-dimensional neural ODE surrogates that accelerate training~\citep{linot2023}. While these reduce cost, their applicability remains configuration-specific and requires careful validation before broad deployment.

In \cite{Viquerat2021}, single-step deep reinforcement learning (SDRL) was presented as a computationally economical variant of DRL, built around a one-step-per-episode interaction scheme. The SDRL is especially effective when the target policy can be treated as state-independent, because it admits compact network representations and a simpler learning problem. This is particularly appealing in flow-control applications, where each interaction with the environment (e.g., a high-fidelity numerical solution) carries a high computational cost.
 Early applications focused on shape and parameter optimization~\citep{Viquerat2021, Viquerat2022, Ghraieb2022, wang2023}, with extensions to open-loop control formulations~\citep{Ghraieb2021}, suggesting a practical pathway toward learning-based controllers with substantially lower computational overhead than full DRL.

Based on previous numerical investigations~\citep{mohammadikalakoo_canada}, where SDRL effectively suppressed TS waves in the frequency domain, in a recent study \cite{Babak2025} introduced a real-time, model-free controller based on SDRL and demonstrated the suppression of Tollmien-Schlichting waves within a purely numerical framework. The key novelty of such an approach is a state-independent policy: instead of reconstructing the evolving flow state, the agent parameterizes a causal FIR mapping that converts upstream reference wall pressure into actuator commands, so temporal flow physics are carried by the signals and filter rather than encoded in the policy itself. This yields amplitude-phase opposition with compact networks, fast convergence, and strong attenuation for single- and multi-frequency content, while remaining robust to uncorrelated sensor/actuator noise. Benchmarks against the classical FXLMS controller indicated superior performance with quicker adaptation, achieved without an explicit plant model. 

Building on this foundation, it is important to note that SDRL-based controllers for TS wave suppression have so far been demonstrated primarily in numerical settings. While some of these studies~\citep{Babak2025} explicitly considered idealized noise sources (e.g. synthetic additive sensor/actuator noise or uncorrelated disturbances), they remain non-empirical and therefore do not fully capture several key challenges that emerge in wind-tunnel operation. In an experimental setting, the controller must learn in real-time under imperfect sensing (microphone noise, calibration drift), uncertain and velocity-dependent convection delays, finite sampling and computation latency, and strong electromagnetic interference generated by DBD actuation. Moreover, the specific type of actuator architecture (i.e. surface DBD)  introduces hard voltage limits, amplitude saturation, and nonlinear control authority as well as potential performance drift due to thermal/electrical loading, which collectively constrain exploration and can compromise learning stability if not handled carefully.

Motivated by this gap, the present study implements the SDRL-based controller in an experimental wind-tunnel configuration using DBD plasma actuation and wall-pressure sensing to assess feasibility, stability, and efficiency under realistic experimental constraints. The problem is formulated in a canonical feedforward layout with an upstream reference microphone, a downstream error microphone, and a spanwise-uniform DBD actuator for control. Controller convergence, learned filter characteristics, and disturbance attenuation are evaluated across progressively more challenging excitation environments (single-frequency, multi-frequency, and broadband forcing), and robustness to moderate variations in freestream velocity is examined. Finally, the downstream persistence of attenuation is assessed using additional microphones and diagnosed with the two-component planar particle image velocimetry~(PIV).

The paper is organized as follows. Section~\ref{Sec:methodology} describes the experimental methodology, including the wind-tunnel facility, flat-plate model, sensing and actuation layout, and data acquisition. Section~\ref{Sec: T and C} introduces the TS wave generation and summarizes the SDRL formulation and implementation. Section~\ref{Sec: Results} presents the wind-tunnel results for single-frequency, multi-frequency, and broadband excitations, covering controller convergence, learned filter characteristics, spectral attenuation, downstream persistence of attenuation, and PIV diagnostics. The effect of freestream velocity is discussed separately in Appendix~\ref{App:freestream}. Finally, Section~\ref{Sec: conclusion} concludes the paper by summarizing the main findings, discussing implications for transition delay, and outlining directions for future research.

\section{Methodology} \label{Sec:methodology}
\subsection{Problem definition}

The experimental implementation and performance assessment of the SDRL-based controller are conducted in a wind tunnel facility under low-turbulence, zero-pressure-gradient conditions. The objective is to attenuate convecting TS waves within a flat-plate boundary layer using a model-free, real-time control framework. Preliminary analyses based on the incompressible Blasius boundary layer solution and the Orr-Sommerfeld stability equation are used to define a range for the operating parameters and determine the spatial placement of sensors and actuators following~\cite{Babak2025}.

\begin{figure}
    \centering
    \includegraphics{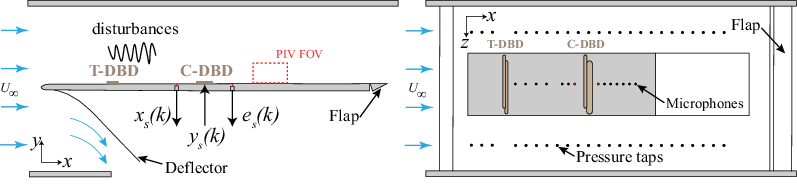}
    \captionsetup{justification=centering}
    \caption{Side and top views of the experimental setup. The signals $\bm{x_s}(k)$, $\bm{y_s}(k)$, and $\bm{e_s}(k)$ denote the reference, actuator, and error signals, respectively. ``T-DBD'' and ``C-DBD'' indicate the plasma actuators used for triggering and controlling the TS waves, respectively. A deflector is installed beneath the flat plate to shield the microphones from the incoming flow. The region used for PIV measurements is outlined by a dashed red box~(schematic not to scale).}

    \label{fig:flow_problem}
\end{figure}

The experimental setup, illustrated in Figure~\ref{fig:flow_problem}, consists of a modular flat plate instrumented with two spanwise-uniform DBD plasma actuators and a linear array of fourteen flush-mounted microphones. The upstream actuator (T-DBD) is used to generate TS wave disturbances, while the downstream actuator (C-DBD) is driven by the SDRL controller to suppress them. It must be noted that the signal sent to actuator T-DBD is entirely independent of the sensing and control loop. The excitation cases are defined in Section~\ref{Sec: T-DBD}.

Throughout this work, spatial quantities are non-dimensionalized using characteristic scales of the baseline Blasius boundary layer. Specifically, streamwise and wall-normal coordinates are scaled by the local displacement thickness ($\delta^{*}_{\mathrm{ref}} = 1.29~\mathrm{mm}$) at the reference location of the C-DBD controller actuator~($x = 0$). Velocities are normalized by the freestream velocity~($U_\infty = 20~\text{m/s}$). Time is non-dimensionalized using the convective time scale $\delta^{*}_{\mathrm{ref}}/U_\infty$, such that $t = t_{\mathrm{dim}} U_\infty/\delta^{*}_{\mathrm{ref}}$. The non-dimensional frequency is expressed by $F = \omega_{ref}/Re_{\delta^{*},\mathrm{ref}} \times 10^6$, where $\omega_{ref} = 2\pi f \delta^{*}_{\mathrm{ref}}/U_\infty$ and $f$ is the dimensional frequency. Using these reference scales, the global Reynolds number is defined as $Re_{\delta^{*},\mathrm{ref}} = U_\infty \delta^{*}_{\mathrm{ref}}/\nu \approx 1.7\times10^{3}$, where $\nu$ is the kinematic viscosity. The coordinate system is defined such that $x=0$ corresponds to the controller actuator center, $y=0$ to the wall surface, and $z=0$ to the midspan of the test section.

The control objective is to synthesize, in real time, an actuation command~($\bm{y_s}$) that minimizes the downstream disturbance amplitude~($\bm{e_s}$) using only the upstream microphone signal~($\bm{x_s}$), subject to practical sensing and actuation constraints such as voltage limits and control stability.

\subsection{Experimental setup and flow conditions}

Experiments are carried out in the anechoic A-Tunnel (open-jet) facility at Delft University of Technology. The test section has a rectangular cross-section~$(250\times860~\mathrm{mm}^2)$ and is acoustically treated to provide a low-noise environment suitable for instability measurements. The turbulence intensity remains below $0.05\%$, and the outlet velocity uniformity is within $\pm0.6\%$ of $U_\infty$ for the operating range of velocities~\citep{MerinoMartinez2020}.

The wind tunnel model shown in Figure~\ref{fig:flow_problem} consists of a modular aluminum flat plate measuring $860~\mathrm{mm}$ in width, $1350~\mathrm{mm}$ in length, and $10~\mathrm{mm}$ in thickness, designed for transition and flow-control studies. A super-elliptic leading edge ensures a smooth pressure-gradient evolution that minimizes flow separation and prevents early transition, thereby maintaining a stable laminar boundary layer essential for transition-delay experiments~\citep{Lin1992}. The overall pressure distribution can also be fine-tuned by varying the angle of the trailing-edge flap, providing additional control over the streamwise pressure gradient. The wall-pressure field is verified using two staggered rows of 128 pressure taps distributed along the span at~$z=\pm248$, to monitor both streamwise pressure gradient and spanwise pressure uniformity. Some of these taps are shown as black dots in Figure~\ref{fig:flow_problem} (top view).

A two-layer poly(methyl methacrylate) (PMMA; Plexiglas) insert ($399\times719~\mathrm{mm}^2$) is mounted flush with the plate surface and accommodates both plasma actuators and the microphone array. The dielectric layer is~$1~\mathrm{mm}$ thick and bonded to a~$8~\mathrm{mm}$ PMMA substrate, providing a total thickness of approximately~$9.8~\mathrm{mm}$ after adjustment by plain washers. The upstream (T-DBD) and downstream (C-DBD) actuators are positioned at~$x=-279$~($380~\mathrm{mm}$) and $x=0$~($740~\mathrm{mm}$) from the leading edge, respectively, as shown in Figure~\ref{fig:flow_problem}. This configuration ensures that the control actuator interacts with disturbances within the region of maximum TS wave amplification while maintaining a laminar base flow. A detailed discussion on the positioning of actuators is provided in Section~\ref{sec:flow and stability}.

\subsubsection{Microphone measurements}

Surface pressure fluctuations are monitored using miniature electret condenser microphones (PUI Audio POM-2739L-HD3-LW100-R, sensitivity $14.0~\mathrm{mV/Pa}$, nominal frequency range 20~Hz–20~kHz). Microphone outputs are reported as voltage normalized by a nominal reference value $V_{\mathrm{ref}}=1~\mathrm{V}$ (i.e.\ $V/V_{\mathrm{ref}}$), yielding a dimensionless amplitude.
Each microphone is mounted flush with the upper surface of the flat-plate model through a $0.4~\mathrm{mm}$ diameter orifice connected to a $6.1~\mathrm{mm}$ cavity machined in the substrate. The cavity is sealed with epoxy to ensure acoustic isolation from the ambient environment and to prevent leakage through the mounting holes. The resulting Helmholtz-type resonance frequency of the combined tap–cavity system is of $\mathcal{O}(\mathrm{kHz})$, well above the frequency band of interest ($150–350~\mathrm{Hz}$) and therefore does not interfere with the measured TS wave content.

A total of fourteen microphones are distributed along the streamwise centerline of the plate as shown with black dots in Figure \ref{fig:flow_problem}.
The array covers the region from $x=-380$ to $x=155$, with a spacing of $19$~($25~\mathrm{mm}$) in the control region and $39$~($50~\mathrm{mm}$) elsewhere. 
Two microphones were assigned as the reference and error sensors in the feedforward control configuration, positioned approximately $x=\pm58$~($75~\mathrm{mm}$) upstream and downstream of the control actuator, respectively, to avoid transient effects induced by the plasma-generated wall jet. This configuration enables a direct comparison between the incoming and attenuated disturbance levels.

To prevent contamination from the plasma discharge, all microphones are embedded beneath a $1~\mathrm{mm}$ dielectric layer with their ports flush to the surface, minimizing local flow disturbance. 
The wiring and signal conditioning are housed below the plate to reduce electromagnetic interference from the high-voltage DBD operation. Signals are amplified and digitized at a sampling frequency of~$5.6~\mathrm{kHz}$.

\subsubsection{Two-component velocity field measurements}\label{Sec:PIV-methodology}

Two-component particle image velocimetry (PIV) is employed to measure the in-plane velocity field ($u,v$) within the streamwise–wall-normal ($x,y$) mid-plane and to characterize the evolution of the TS waves with and without control. The laser sheet is aligned slightly offset to the test-section midspan to minimize surface reflections from the microphone orifices. The measurement field of view (FOV) spans~$35 \times 8$, centered approximately at $x=97$ downstream of the control actuator shown by dashed boxes in Figure~\ref{fig:flow_problem} (not to scale). This area is selected to capture the evolution and growth of the TS waves past the controller actuator while maintaining sufficient spatial resolution near the wall, where velocity gradients are steep.

For all case studies, the flow is seeded with tracer particles of approximately $1~\mu\text{m}$ mean diameter, generated by a fog generator using a water–glycol mixture. The illumination plane is formed by a dual-cavity Nd:YAG~(neodymium-doped yttrium aluminum garnet) laser (200~mJ per pulse), which produces a laser sheet of about~$1~\mathrm{mm}$ thickness through a combination of cylindrical and spherical lenses.

\begin{table}
\centering
\caption{Specifications of the PIV setup.}
\label{Table:PIV}
\renewcommand{\arraystretch}{1.1}
\setlength{\tabcolsep}{6pt}
\begin{tabular}{l c}
\toprule
\toprule
\text{Parameter} & \text{Specification} \\
\midrule
Sensor resolution & $2560 \times 2160~\text{pixels}^2$ \\
Pixel pitch ($\mu$m) & 6.5 \\
Lens focal length (mm) & 200 \\
Frame separation ($\Delta t$) ($\mu$s) & 12 \\
Magnification factor & 0.37 \\
Field of view area (mm$^2$) & $45 \times 10$ \\
Overlap (\%) & 50 \\
Vectors per field & $410 \times 90$ \\
Vector pitch (mm) & $0.11$ \\
\end{tabular}
\end{table}

\FloatBarrier

Image acquisition is performed using a LaVision Imager sCMOS camera with corresponding specifications presented in Table~\ref{Table:PIV}. The inter-pulse separation is set to $\Delta t = 12~\mu\text{s}$, yielding particle image displacements of approximately $10$ pixels in the near-wall region. A total of 2000 image pairs are recorded for each configuration, providing sufficient statistical convergence for estimating mean and fluctuating quantities along the streamwise direction. Image pairs are recorded at a sampling rate of~$13.13~\text{Hz}$ under non-time-resolved operation to avoid phase locking with the periodic TS wave excitation frequencies. This ensures that each captured image pair corresponds to a different phase of the oscillatory flow, yielding statistically independent velocity fields for accurate ensemble-averaged analysis. 

Velocity fields are computed using multi-pass cross-correlation with decreasing interrogation-window sizes, starting from $64\times64$ pixels and refined to $12\times12$ pixels with 50\% overlap. The resulting vector fields are further processed to remove spurious vectors through local median filtering and validation criteria based on signal-to-noise ratio thresholds. The PIV results are presented in Section~\ref{sec:piv_validation}, where the influence of control on the boundary layer structure and TS wave attenuation is analyzed.

The uncertainty in the time-averaged velocity fields was estimated by correlation statistics~\citep{wieneke2015piv, sciacchitano2019uncertainty}, and the corresponding maximum uncertainties in the velocity components were $\varepsilon_u = 0.08\%~U_\infty$ and $\varepsilon_v = 0.05\%~U_\infty$.

\subsubsection{Flow condition characteristics}\label{sec:flow and stability}

\begin{figure}
    \centering
    \includegraphics{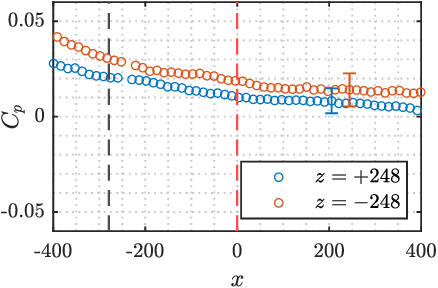}
    \captionsetup{justification=centering}
    \caption{Pressure coefficient distribution for two arrays of pressure taps shown in Figure~\ref{fig:flow_problem}. The locations of the T-DBD and C-DBD actuators~(black and red dashed lines). The error bars indicate the standard deviation of the measurements.}
   \label{fig:p_gradient}
\end{figure}

The characterization of the undisturbed flat-plate boundary layer is carried out in the absence of TS wave excitation, with both actuators inactive. The pressure coefficient distribution, $C_p = (p-p_{\infty})/(p_0-p_{\infty})$, recorded by two symmetric arrays of pressure taps, is presented in Figure~\ref{fig:p_gradient}. As indicated in Figure~\ref{fig:flow_problem}, the taps are positioned on either side of the midspan at $z=\pm{}248$. The close agreement between the two pressure distributions confirms that the boundary layer in the measurement region can be considered two-dimensional within the experimental uncertainty. The streamwise pressure-gradient level was quantified using the acceleration parameter
$K = (\nu/U_e^{2})\,\mathrm{d}U_e/\mathrm{d}x$~\citep{schultz2007rough}, and found to remain well below the zero-pressure-gradient criterion in the region $-400 \le x \le 400$, with $\max |K| = 3.269 \times 10^{-8}$ and $\mathrm{RMS}(K) = 1.253 \times 10^{-8}$ (threshold $|K| < 1.6 \times 10^{-7}$).


\begin{figure}
    \centering
    \includegraphics{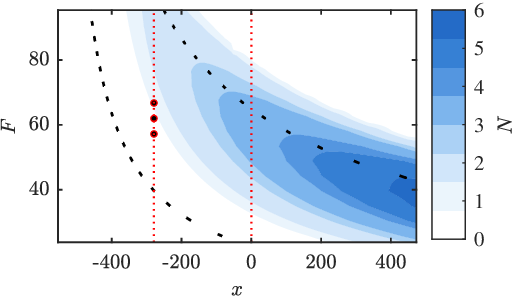}
    \captionsetup{justification=centering}
   \caption{Linear stability diagram at $Re_{\delta^{*},\mathrm{ref}} \approx 1.7\times10^{3}$~($U_\infty = 20~\mathrm{m/s}$), showing the amplification factor ($N$) as blue contours. The neutral stability curve ($\alpha_i=0$)~(black dashed lines). The T-DBD and C-DBD at $x=-279$ and $x=0$, respectively~(red dotted lines). The TS wave excitation frequencies triggered by the T-DBD actuator~(bold black markers).}

   \label{fig:stability_curve}
\end{figure}

The stability characteristics of the laminar boundary layer are provided in Figure~\ref{fig:stability_curve} using spatial linear stability theory~(LST) based on the incompressible Orr-Sommerfeld equation~\citep{Mack84}, assuming a zero-pressure-gradient Blasius base flow (i.e.~$C_p=0$). The amplification factor of TS waves is computed for a range of excitation frequencies, $F = 0-238$  ($0-1000~\mathrm{Hz}$), to identify the frequency band associated with the most amplified disturbances along the streamwise direction.

The boundary conditions impose zero velocity perturbations ($u' = v' = 0$) at both the wall and the freestream. The resulting neutral stability curve ($\alpha_i = 0$), bold dashed line in Figure~\ref{fig:stability_curve}, exhibits the classical behavior of a Blasius boundary layer, providing a reference for selecting the experimental actuation frequencies. The triggered TS wave frequencies used in the experiments are chosen within the linearly unstable regime, ensuring measurable amplification without triggering nonlinear effects or early transition as shown in Figure \ref{fig:stability_curve}. The amplification factor, or $N$-factor, shown with blue contours in Figure~\ref{fig:stability_curve}, is obtained by integrating the negative imaginary part of the complex wavenumber, $-\alpha_i(x,f)$, from the onset of instability $x_0$ (where $\alpha_i = 0$) to a given downstream location $x$, as

\begin{equation}
N(x,f) = \int_{x_0}^{x} -\alpha_i(x',f)\,\mathrm{d}x',
\label{eq:n_factor}
\end{equation}
which represents the cumulative spatial growth of each frequency mode and quantifies the amplification of small disturbances as they convect downstream. The resulting $N$-factor map, $N(x,f)$, provides a clear visualization of the instability region and identifies the frequency range around $F\approx60$ ($f \approx 250~\mathrm{Hz}$) as the most amplified under the present flow conditions. Based on this analysis, the actuator locations are selected within the frequency band where the laminar boundary layer exhibits pronounced TS wave amplification, approximately~$F = 36-83$~($150-350~\mathrm{Hz}$), with the dominant amplification peak near $F = 60$ close to the control position. Accordingly, representative trigger signals, shown with black dots in Figure~\ref{fig:stability_curve}, are defined as a single-frequency excitation at $F=57$~($f = 240~\text{Hz}$), a two-tone excitation at~$F = [57,62]$~($f = [240,260]~\text{Hz}$), a three-tone excitation at~$F=[57,62,67]$~($f = [240,260,280]~\text{Hz}$), and broadband white-noise forcing, covering the range of disturbance environments explored in the control experiments.

\section{TS wave excitation and control framework}\label{Sec: T and C}

This section describes the elements required to generate TS wave disturbances and to implement the real-time SDRL-based suppression strategy in the wind tunnel. First, the upstream trigger actuator and the resulting excitation cases are defined to prescribe the incoming disturbance spectrum. Then, the downstream control actuator and the full experimental control framework are detailed, including signal conditioning, synchronization, and the real-time actuation chain.

\subsection{TS wave generation and test cases} \label{Sec: T-DBD}

Tollmien-Schlichting waves are generated using a spanwise two-dimensional DBD plasma actuator~\citep{Marios2015}, referred to as T-DBD, which is integrated into a custom insertion plate. This actuator operates under alternating-current (AC) excitation supplied by TREK~20/20C amplifiers. The actuator assembly consists of a PMMA substrate and dielectric layers, with thicknesses of $8~\mathrm{mm}$ and $1~\mathrm{mm}$, respectively. The electrode consists of self-adhesive copper tape with a thickness of approximately $50~\mu\text{m}$, nearly an order of magnitude smaller than the local displacement thickness (on the order of $~\mathrm{mm}$). Hence, the surface roughness introduced by the actuator is considered negligible. A streamwise overlap of $-2~\mathrm{mm}$ between the exposed and covered electrodes is implemented to ensure uniform spanwise plasma formation. 

In this study, direct sinusoidal modulation is employed to generate a stable plasma with reduced electromagnetic noise and slower actuator degradation, offering a longer operational lifetime compared to high-frequency carrier signal modulation, while still providing sufficient strength to test the controller. The maximum peak-to-peak control voltage is $20~\text{kV}$, although the system is capable of operating up to $40~\text{kV}$. 

Three excitation scenarios were designed to progressively increase the complexity of the control task. The single-frequency case~$A$,~$F = 57 $, provides a deterministic, single-tone disturbance that is used to validate the setup and establish baseline controller performance. The multi-frequency case~$B1$,~$F=[57,62]$, increases complexity by superimposing two modes with different frequencies and random initial phases, producing a time-varying waveform and placing higher demands on phase and amplitude matching. The three-tone case~$B2$,~$F=[57,62,67]$, further increases spectral richness and waveform irregularity, bridging deterministic and broadband forcing. Finally, the broadband white-noise cases~$C1$ and $C2$ represent broadband excitation with random phase and amplitude content at two freestream velocities,~$Re_{\delta^{*},\mathrm{ref}} \approx 1700$~($U_{\infty} = 20~\text{m/s}$)~and~$Re_{\delta^{*},\mathrm{ref}} \approx 1912$~($U_{\infty} = 22.5~\text{m/s}$), which provide the most challenging studied configuration for assessing robustness and performance of the SDRL controller.

\begin{figure}
  \centering

  \begin{minipage}[t]{0.56\textwidth}
    \centering
    \includegraphics[width=\linewidth,trim=1 0 0 0,clip]{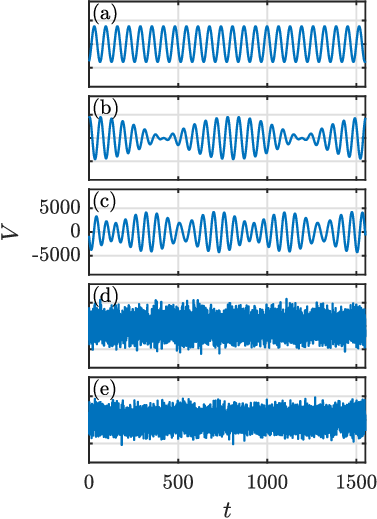}
  \end{minipage}
  \hspace{0.015\textwidth} 
  \begin{minipage}[t]{0.40\textwidth}
    \centering
    \includegraphics[width=\linewidth,trim=0 0 0 0,clip]{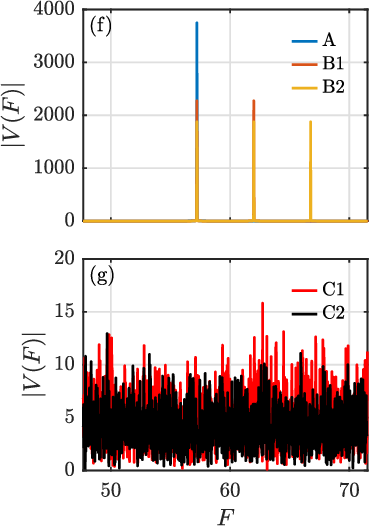}
  \end{minipage}

  \caption{T-DBD voltage signals ($a-e$) and their fast Fourier transform (FFT) ($f-g$) for cases: ($a$) $A$, ($b$) $B1$, ($c$) $B2$, ($d$)~$C1$, ($e$) $C2$.}
  \label{fig:trigger_voltage_combined}
\end{figure}

Figure~\ref{fig:trigger_voltage_combined} illustrates the voltage signal applied to the triggering actuator and its corresponding frequency spectrum for representative cases~$A$, $B1$, and~$C1$. The progressive broadening of the frequency content—from a single sharp peak in case~$A$, to two discrete tones in case~$B1$, and finally a continuous broadband distribution in case~$C1$—clearly reflects the intended escalation in excitation complexity. The main experimental parameters for all test cases, including excitation frequencies, control voltage ranges, and sampling settings, are summarized in Table~\ref{Table:control_cases}, overviewing the operating conditions.

\begin{table}
\centering
\caption{Experimental parameters for all control cases. The definitions and detailed justification of the control-related parameters (e.g., FIR filter length, action bound, sampling rate, and RMS window) are provided in Section~\ref{Sec:exp_control}.}
\label{Table:control_cases}
\begin{tabular}{lccccc}
\toprule
\toprule
\text{Parameter} 
& \text{$A$} 
& \text{$B1$} 
& \text{$B2$} 
& \text{$C1$} 
& \text{$C2$} \\
\midrule
Description & Single-freq. & Multi-freq. & Multi-freq. & White-noise & White-noise \\
$Re_{\delta^{*},\mathrm{ref}}~(U_{\infty}~[m/s])$ & 1700~(20) & 1700~(20) & 1912~(22.5) & 1700~(20) & 1912~(22.5) \\
FIR filter coefficients & 40 & 60 & 60 & 60 & 60 \\
Action bound & $\pm10$ & $\pm7$ & $\pm7$ & $\pm7$ & $\pm7$ \\
Max control voltage~[kV] & 15 & 17 & 17 & 15 & 15 \\
Sampling rate~[kHz] & 5.0 & 5.0 & 5.6 & 5.6 & 5.6 \\
RMS window & 10~periods & 40~periods & 40~periods & 40~periods & 40~periods \\
Trigger frequency & 57 & [57,\,62] & [57,\,62,\,67] & broadband & broadband \\
Trigger voltage~[kV] & 7.5 & 9.1 & 8.5 & 9.5 & 8.0 \\

\end{tabular}
\end{table}

\FloatBarrier

\subsection{Controller actuator}

The downstream actuator, referred to as the control DBD (C-DBD), located at $x = 0$, shares the same structural configuration and materials as the upstream T-DBD actuator described in Section~\ref{Sec: T-DBD}. The upper (powered) electrode is the only element exposed to the external flow and slightly protrudes above the flat-plate surface by about $50~\mu\text{m}$, while the grounded electrode and interconnections are fully embedded within the laminate. This design minimizes the geometric step and ensures a nearly smooth aerodynamic surface, such that the boundary layer remains practically undisturbed—an essential feature for transition-control experiments.

The ground electrode has a streamwise width~$\approx19~(25~\mathrm{mm})$, increased relative to the trigger actuator~$\approx8~(10~\mathrm{mm})$ as shown in Figure~\ref{fig:flow_problem}, to prevent plasma constriction and maintain control authority at higher voltages. The ground electrodes are insulated by a thin Kapton layer to avoid unwanted discharges within the laminate.

The actuator is powered by a TREK~20/20C HS high-voltage and high-speed amplifier driven directly by the real-time controller during closed-loop experiments and by a digital function generator in open-loop tests. During the experimental campaign, sinusoidal excitation with peak-to-peak amplitudes up to $20~\text{kV}$ was applied. The controller output signal computed in real time from the reference microphone measurements is filtered and soft-clipped before amplification to minimize electromagnetic interference and actuator degradation. Further details on the actuator signal and control framework are provided in Section~\ref{Sec:exp_control}.

\subsection{Control framework}\label{Sec:exp_control}

\begin{figure}
    \centering
    \includegraphics{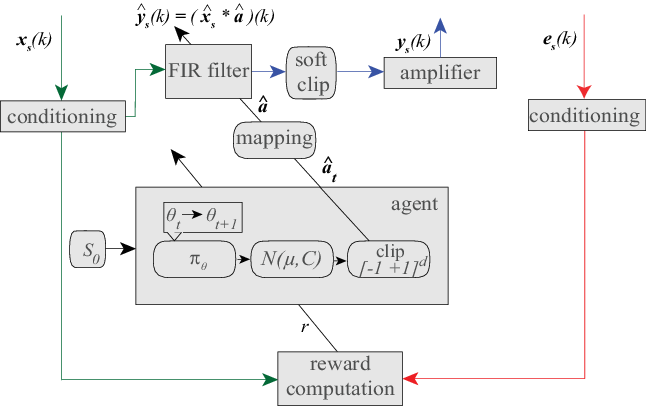}
    \captionsetup{justification=centering}
    \caption{Schematic of the control unit. $\bm{x_s}(k)$, $\bm{y_s}(k)$, and $\bm{e_s}(k)$ refer to reference, actuator, and error signals, respectively. Figure is adjusted from~\cite{Babak2025}.}
   \label{fig:controller_flowchart}
\end{figure}

The control architecture follows the methodology introduced by~\cite{Babak2025}, where the SDRL-based controller for TS wave attenuation was assessed numerically. The present work follows an identical control procedure comprising reference-based FIR filtering, reward formulation, and SDRL updates, but extends it to a real-time experimental setup in a wind tunnel, allowing a feasibility check of this type of controller. The flowchart of the controller framework is presented in Figure \ref{fig:controller_flowchart}.

The control framework couples the SDRL agent with the physical system through a real-time simulation and testing platform. The framework performs signal acquisition, digital filtering, reward evaluation, and high-voltage actuation synthesis. Reference ($\bm{x_s}$) and error ($\bm{e_s}$) microphone signals are conditioned through a sequence of digital filters: an initial low-pass filter removes broadband noise, followed by a band-pass filter ($150-350~\mathrm{Hz}$) that isolates the TS wave content for accurate reward ($r$) computation. The instantaneous wave amplitude is determined from the root-mean-square (RMS) value of the filtered signals. A reference-error synchronization module compensates for the convection delay between microphones, ensuring precise temporal alignment for control performance evaluation and computation of the reward. The reward computation and the details on the SDRL algorithm are provided in Section~\ref{Sec: SDRL}.

The synthesized control signal as an output of the SDRL algorithm is subsequently low-pass filtered and passed through a soft-clipping function to prevent amplifier overshoot and minimize actuator degradation. The resulting waveform is amplified by a TREK~20/20C high-voltage amplifier and applied to the C-DBD actuator in real time. 

The control parameters for each case are shown in Table~\ref{Table:control_cases}, which are guided by the spectral complexity of the excitation and the stability requirements of the hardware–software interface. The action bound in Table~\ref{Table:control_cases} defines the allowable range for the controller’s output during learning, limiting the magnitude of the FIR filter coefficients and thus the actuation voltage. A wider bound (e.g., $\pm 10$) allows broader exploration (including more aggressive control actions) when the hardware-software loop remains stable, whereas a tighter bound (e.g., $\pm 7$) is used in scenarios that are more prone to nonlinear responses and saturation to preserve stability during learning. The action bound constrains the controller’s numerical output during learning, while the maximum control voltage represents the physical hardware limit of the plasma actuator.

In case~$A$~(single-frequency), a 40-coefficient FIR filter was sufficient to represent the quasi-linear disturbance at $F = 57$, while the wider amplitude range ($\pm10$) allowed the controller to explore stronger actuation magnitudes without saturating the voltage signal. As more frequency components were introduced in cases $B1$ and $B2$, the FIR filter length was extended to 60 coefficients to provide higher temporal resolution and capture phase interactions among modes. The action bound was consequently reduced to $\pm7$ to maintain learning stability and prevent amplifier saturation, since longer filters could produce stronger signal peaks.

In $B1$ (two tones at $[57,\,62]$ and $Re_{\delta^{*},\mathrm{ref}}\approx1700$), the trigger voltage was increased to $9.1~\text{kV}$ to ensure clean multi-tone excitation and compensate for the reduced amplitude caused by modal beating. The control voltage cap was raised to $17~\text{kV}$ to sustain effective plasma actuation within the broader frequency band. In $B2$ (three tones at $[57,\,62,\,67]$ and $Re_{\delta^{*},\mathrm{ref}}\approx1912$), the higher freestream velocity enhanced disturbance amplification, allowing a lower trigger voltage $8.5~\text{kV}$ to achieve comparable wave amplitudes. The same FIR filter length and control voltage cap were retained, while the sampling rate was increased to $5.6~\text{kHz}$ to accurately resolve the additional higher-frequency component. 

In case $C1$, broadband excitation required higher trigger voltages up to $9.5~\text{kV}$ to maintain sufficient energy across the wide frequency spectrum. However, the control voltage was limited to $15~\text{kV}$ since actuation energy is inherently distributed across multiple frequencies~$F = 36-83$~($150-350~\mathrm{Hz}$), and higher voltages would impose unnecessary electrical stress on the actuator. 

Overall, these parameter variations reflect a balance between controller adaptability and actuator safety, ensuring stable learning performance as the control problem evolved from deterministic single-frequency disturbances to broadband, stochastic excitation environments.

\subsubsection{SDRL-based controller}\label{Sec: SDRL}

For the SDRL agent, the control waveform is synthesized as the convolution of the reference signal~($\bm{x_s}$) with an FIR filter whose coefficients form the policy vector~($\bm{\hat{a}}$). These coefficients are iteratively optimized by the SDRL algorithm to maximize the reward~($r$) derived from the RMS of the filtered error signal~($\bm{e_s}$). The SDRL algorithm updates the filter coefficients by sampling them from a parameterized multivariate normal distribution, $\pi_{\theta}(\bm{\hat{a}}) = \mathcal{N}(\bm{\mu},\bm{C})$, whose parameters $\bm{\theta} = {\bm{\mu},\bm{C}}$ are adjusted through gradient ascent on the expected reward. At the beginning of each generation, the agent samples a population of candidate actions~($\bm{\hat{a}}_t$) (i.e., sets of FIR coefficients) from this distribution and evaluates their corresponding rewards. The distribution parameters are then updated so that future samples are biased toward actions yielding higher rewards. In contrast to value-based or multi-step DRL schemes, the SDRL approach treats each generation as an independent optimization problem, thereby eliminating the need for temporal backpropagation through a sequence of states and allowing deterministic, low-latency updates. For a comprehensive explanation of the algorithm’s implementation details and notation, the reader is referred to~\cite{Viquerat2022} for the SDRL algorithm and~\cite{Babak2025} for the SDRL-based controller framework.

\FloatBarrier
\begin{table}
\centering
\caption{Details of the SDRL hyperparameters.}
\begin{tabular}{ll}
\toprule
\toprule
\text{Parameter} & \text{Value / Setting} \\
\midrule
Neural network architecture & [2, 2, 2] (hidden layers per network) \\
Initial learning rates & $5\times10^{-4}$ for $\mu, \sigma^{2}$; $1\times10^{-4}$ for $\rho$ \\
Number of individuals per generation & $10$ \\
Total generations & $150-500$ \\
Decay factor & $0.98$ \\
\bottomrule
\label{Table: hyper_param}
\end{tabular}
\end{table}

A distinctive feature of SDRL is that the policy network receives a constant input state, $S_0$, rather than a time-varying observation. Temporal information relevant to the control problem is embedded in the reference signal~($\bm{x_s}$) and the FIR filter itself, which collectively encode the convective delay and spectral content of the incoming TS waves. Consequently, the agent learns a stationary mapping between the probability distribution of FIR coefficients and the attenuation reward, effectively shaping the control waveform in both amplitude and phase without explicitly modeling time evolution. This compact formulation drastically reduces network complexity and computational burden, making the approach suitable for high-frequency real-time control.  A summary of the SDRL hyperparameters used in this study is presented in Table \ref{Table: hyper_param}. Values are based on the configuration of~\cite{Babak2025} and adapted as a trade-off between stability, exploration range, and actuation strength—sufficient to capture the temporal window of the TS wave packet while ensuring robust training and physically meaningful voltage levels at the actuator.

An adaptive learning-rate scheme was implemented within the SDRL algorithm to enhance convergence stability and efficiency. Each of the three neural 
networks defining the policy, those corresponding to the mean ($\mu$), variance 
($\sigma^2$), and correlation ($\rho$) parameters of the multivariate Gaussian distribution, was 
assigned an independent learning rate ($\mathrm{lr}_\mu$, $\mathrm{lr}_{\sigma^2}$, $\mathrm{lr}_\rho$). The algorithm 
monitored the average reward over a moving window of 
$20$ generations and compared it with the previous window to detect 
stagnation in performance. When the relative improvement in reward was below 
$2~\%$, the learning rates were reduced by a factor of $k_\mu = k_{\sigma^2} = k_\rho = 0.5$, 
while ensuring that they did not fall below their respective minimum values 
($\mathrm{lr}_{\mu,\min} = \mathrm{lr}_{\sigma^2,\min} = \mathrm{lr}_{\rho,\min} = 10^{-5}$). 
This mechanism allowed large learning rates during the early exploratory phase of training 
and progressively smaller values as the policy approached convergence, preventing 
oscillations once the controller achieved steady attenuation performance. In practice, the 
initial learning rates were set according to Table \ref{Table: hyper_param}. The adaptive schedule was activated after 
$100$ generations, which corresponded to the typical onset of reward saturation in preliminary trials.

The reward quantifies the control performance in terms of the attenuation of the TS waves measured by the microphones. It is computed in real time and used by the SDRL agent to update its policy as shown in the flowchart of Figure~\ref{fig:controller_flowchart}. The reference ($\bm{x_s}$) and error ($\bm{e_s}$) microphone 
signals are simultaneously sampled through a data-acquisition module. Both signals are first low-pass filtered to remove broadband acoustic noise and subsequently 
band-pass filtered in the range $150-350~\mathrm{Hz}$ to isolate the TS wave content relevant for control evaluation. 

The instantaneous amplitude of the filtered pressure fluctuations is estimated using a 
sliding RMS window.

\begin{equation}
V_\mathrm{RMS}(n) = 
\sqrt{\frac{1}{N_\mathrm{RMS}} 
\sum_{i=0}^{N_\mathrm{RMS}-1} V^2(n-i)}.
\label{eq:RMS}
\end{equation}
Here $N_\mathrm{RMS}$ represents the number of samples within the RMS window. 
The window length corresponds to approximately $10$ oscillation periods of the lowest dominant frequency (wave periods) for the case~$A$~(single-frequency) and $40$ periods for multi-frequency or broadband cases indicated in Table~\ref{Table:control_cases}, providing a compromise 
between temporal responsiveness and noise suppression. In case~$A$~(single-frequency), the TS waves exhibit a nearly periodic structure, so averaging over $10$ cycles is sufficient 
to yield a stable estimate of their amplitude while maintaining a fast response to any variations introduced by the controller. Using shorter windows would make the reward susceptible to instantaneous fluctuations and acoustic noise, whereas longer averaging intervals would delay the feedback response and slow down policy adaptation. Conversely, in multi-frequency or broadband excitations, the superposition of several modes produces a slowly varying amplitude envelope caused by beating and nonlinear interactions. In such conditions, a longer averaging window of about $40$ periods is necessary to capture the overall energy content of the signal and avoid interpreting transient interference patterns as meaningful variations in control performance. This choice ensures that the reward reflects the sustained attenuation of TS wave energy rather than short-term oscillations or local phase effects.

A temporal delay is then applied to the reference RMS signal to compensate for the 
convection time of the TS waves between the reference and error microphone locations. 
The delay is estimated from the phase velocity of the waves 
($U_\mathrm{TS} \approx 0.3\,U_{\infty}$)~\citep{Mayes2003Boundary-LayerTheory} and the streamwise spacing between microphones~\citep{Babak2025}. The synchronized RMS values are used to compute the reward according to the selected 
objective function. In the standard formulation, the relative attenuation of the RMS pressure perturbations defines the reward as

\begin{equation}
r = \frac{R_\mathrm{RMS} - E_\mathrm{RMS}}{R_\mathrm{RMS}}.
\label{eq:reward}
\end{equation}
Here $R_\mathrm{RMS}$ and $E_\mathrm{RMS}$ denote the RMS amplitudes of the synchronized 
reference~($\bm{x_s}$) and error signals~($\bm{e_s}$) respectively. Alternative logarithmic and error-based formulations 
can also be used to adjust the reward sensitivity under different flow conditions. 
The resulting reward is then fed back to the SDRL agent, which updates the neural network parameters of the policy.

\section{Results} \label{Sec: Results}

The results of this experimental study are presented according to the control strategy and flow conditions. The controller performance is analyzed in terms of TS wave attenuation and convergence speed for single-frequency, multi-frequency, and broadband excitation cases. For each configuration, the discussion includes the learning evolution, filter characteristics, control-signal behavior, and the corresponding streamwise attenuation of TS wave amplitude. An additional subsection examines the downstream persistence of disturbance attenuation and transition delay. A comparative discussion of the influence of freestream velocity on controller performance is deferred to Appendix~\ref{App:freestream}. Throughout the section, complementary PIV diagnostics are used to support the interpretation of the control behavior and disturbance attenuation.

\subsection{Controller convergence and response}

\begin{figure}
    \centering
        \includegraphics{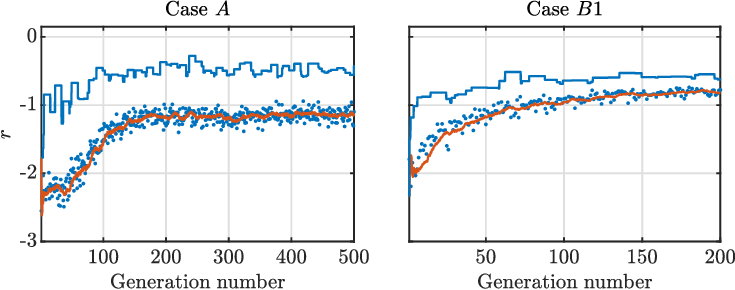}
    \captionsetup{justification=centering}
    \caption{Reward moving average (solid orange line) and moving maximum (solid blue line) evolution during generation of the SDRL algorithm for~$A$ and~$B1$ control cases. The blue dots represent the averaged reward during each generation. The moving window size is 100 episodes (10 generations).}
   \label{fig:reward_conv}
\end{figure}

The convergence behavior of the SDRL-based controller is evaluated through the evolution of the reward during training for both control cases $ A$ (single-frequency) and $B1$ (multi-frequency), as shown in Figure~\ref{fig:reward_conv}. The reward function is defined to increase as the downstream wall-pressure fluctuations decrease, meaning that higher reward values correspond to improved TS wave attenuation. The moving average and maximum of the reward illustrate the temporal evolution of the controller performance over consecutive generations, where each generation represents one optimization cycle of the policy parameters. The moving window size for averaging is 100 episodes, corresponding to 10 generations.

\begin{figure}
    \centering
    \includegraphics{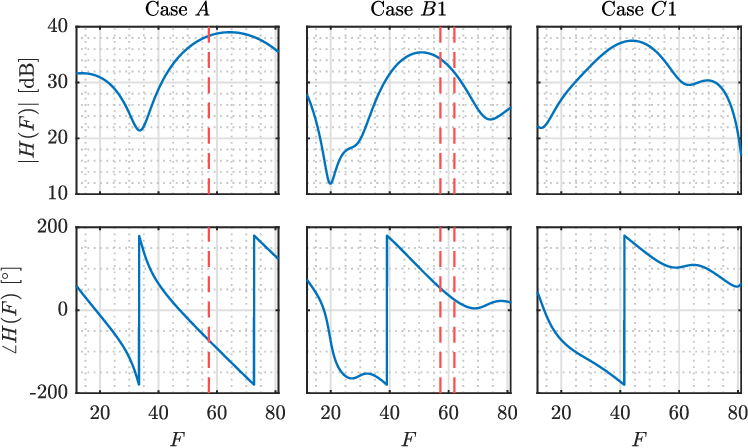}
    \captionsetup{justification=centering}
    \caption{Frequency response of the best-performing FIR filter obtained for $A$, $B1$, and ~$C1$ control cases. The triggered frequencies are shown by dashed red lines in cases~$A$ and~$B1$.}
   \label{fig:filter_response}
\end{figure}

In both control cases, the controller exhibits a rapid rise in reward in less than 100 generations~(1000 interactions with the environment), indicating fast learning of an effective amplitude-phase relation for opposition control. Control time only accounts for the duration of each episode, excluding the communication and generation time, and is therefore independent of the available computational power. After approximately 200 generations, the reward stabilizes near its maximum value, demonstrating convergence toward a consistent policy capable of effectively attenuating the TS disturbances. Minor oscillations in the moving maximum suggest ongoing fine-tuning due to the stochastic exploration inherent to the SDRL algorithm. Overall, these results confirm the robust and autonomous convergence of the SDRL controller, highlighting its capacity to identify effective amplitude-phase actuation policies under varying disturbance conditions. This behavior underscores the algorithm’s adaptability and suitability for real-time flow-control applications, where rapid and stable learning is essential. The reward curves for cases $B2$ and $C1$-$C2$ follow trends similar to those in Figure~\ref{fig:reward_conv} and are therefore omitted for brevity.

\begin{figure}
    \centering
    \includegraphics{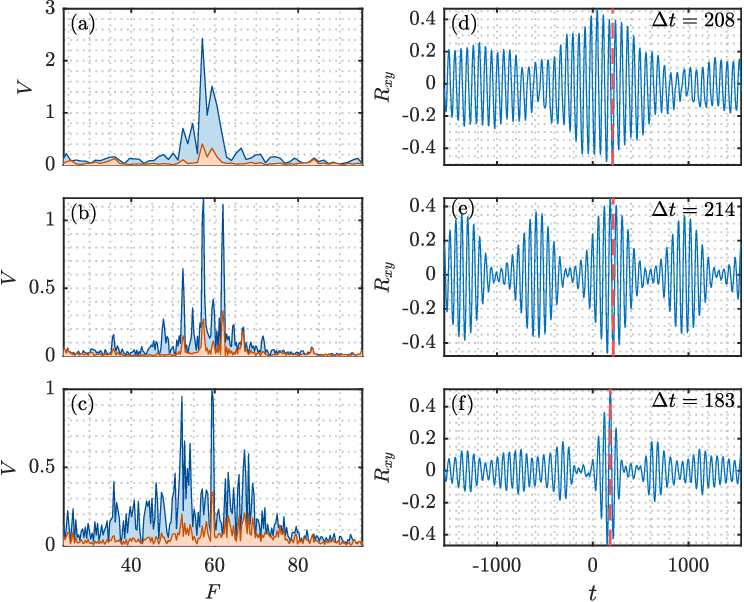}
    \captionsetup{justification=centering}
    \caption{Fourier transform of the plasma controller signal (blue) alongside the amplified reference signal~($\bm{x_s}$) (orange) on ($a-c$), and their cross-correlation on ($d-f$), corresponding to the best-performing FIR filter obtained for~(a,d)~$A$,~(b,e) $B1$, and~(c,f)~$C1$ control cases. Microphone signals are shown as voltage normalized by a nominal reference value $V_{\mathrm{ref}}=1~\mathrm{V}$ (i.e. $V/V_{\mathrm{ref}}$). In all cases, the plasma signal is delayed by $\Delta t$ relative to the reference signal shown by red dashed line.}

   \label{fig:controller_response2}
\end{figure}

The rapid convergence observed in the reward evolution indicates that the SDRL agent has identified a stable reference-to-actuation mapping, represented by the converged FIR filter. Figure~\ref{fig:filter_response} can therefore be interpreted as the learned transfer function $H(F)$ of this mapping. Its phase reflects the delay and phase compensation required to achieve destructive interference downstream, whereas its magnitude indicates how the controller distributes actuation authority across frequency. 

Figure~\ref{fig:controller_response2} provides a complementary signal-level validation of this interpretation. The Fourier spectra in Figures~\ref{fig:controller_response2}$(a$-$c)$ show that the plasma command reproduces the dominant spectral content present in the reference signal~($\bm{x_s}$), indicating that the controller injects actuation energy at physically meaningful TS frequencies rather than generating spurious components. The cross-correlation traces in Figures~\ref{fig:controller_response2}$(d$-$f)$ exhibit a single dominant maximum at a lag $\Delta t$, providing an estimate of the effective delay implemented by the learned FIR mapping, including the combined effect of convection, digital filtering, and actuation-chain dynamics. In other words, the spectra characterize how the controller allocates actuation in frequency, while the cross-correlation quantifies the relative time lag between the reference disturbance and the resulting actuation signal.

In case~$A$~(single-frequency), the controller is primarily required to attenuate a single triggered tone at approximately $F=57~(240~\mathrm{Hz})$. Although Figure~\ref{fig:filter_response} shows that the learned filter exhibits elevated gain over a broader frequency range, this should not be interpreted as genuine broadband control. Since the upstream disturbance is dominated by a single frequency, the optimization primarily constrains the filter response near that tone, while the out-of-band response remains comparatively underdetermined. The Fourier spectrum of the plasma signal in Figure~\ref{fig:controller_response2}(a) supports this interpretation, as it shows a dominant peak at the triggered frequency, matching the reference microphone spectrum. In addition, the phase response of the FIR filter is approximately linear, modulo phase wrapping, which is consistent with a delay-dominated mapping between the reference signal~($\bm{x_s}$) and the actuation command~($\bm{y_s}$).

Using the dominant lag extracted from the cross-correlation in Figure~\ref{fig:controller_response2}(d), the controller voltage is delayed by $\Delta t \approx208~(13.4~\mathrm{ms})$ relative to the reference signal~($\bm{x_s}$), as indicated by the single strong correlation peak. Interpreting this lag as a convective travel time and using an estimated TS wave convective velocity $U_{\mathrm{TS}}\approx 0.3\,U_\infty$ with $U_\infty = 20~\mathrm{m/s}$ yields an effective streamwise distance of~$\Delta x_{\mathrm{eff}}\approx 62$~($80.4~\mathrm{mm}$). This is of the same order as the measured reference-to-actuator spacing of $58$~($75~\mathrm{mm}$). The remaining discrepancy can be attributed to uncertainty in $U_{\mathrm{TS}}$ and additional delays in the sensing-actuation chain. Overall, case~$A$ is therefore best interpreted as a single-tone, delay-and-gain opposition-control solution rather than a truly broadband controller. 


For case~$B1$~(multi-frequency) at~$Re_{\delta^{*},\mathrm{ref}} \approx 1700$, the learned controller must satisfy simultaneous amplitude-phase requirements at multiple triggered frequencies rather than at a single tone. Accordingly, Figure~\ref{fig:filter_response} shows a more structured response around the excitation region than in case~$A$, indicating that the FIR filter is no longer acting as a simple delay-and-gain mapping. Instead, the controller distributes its authority across the multi-tone disturbance band and applies frequency-dependent phase compensation to attenuate the different incoming TS modes simultaneously. This interpretation is consistent with the controller spectrum in Figure~\ref{fig:controller_response2}(b), which contains the dominant spectral components present in the upstream excitation.

Unlike case~$A$, where the nearly linear phase is consistent with a predominantly delay-like mapping, the phase response in case~$B1$ exhibits a more pronounced frequency dependence, reflecting the additional constraints imposed by the simultaneous attenuation of more than one tone. Nevertheless, the cross-correlation in Figure~\ref{fig:controller_response2}(e) still exhibits a single dominant peak at $\Delta t \approx 214~(13.8~\mathrm{ms})$, indicating that the time-domain actuation remains synchronized with the reference disturbance through a dominant effective lag. Case~$B1$ is therefore best interpreted as a multi-tone compensator: more structured and physically constrained than the single-frequency solution in case~$A$, but not yet a genuinely broadband controller in the same sense as the white-noise case.

In case~$C1$~(broadband white-noise), the controller must attenuate TS disturbances distributed over a broad frequency range rather than one or several discrete tones. In this case, the broad response observed in Figure~\ref{fig:filter_response} is physically meaningful, since the incoming disturbance itself contains energy over an extended band. The learned FIR filter therefore behaves as a genuinely broadband compensator: its magnitude remains elevated over a wide range of frequencies, while its phase varies smoothly, indicating frequency-dependent compensation across the unstable TS band rather than single-tone or multi-tone tuning. This interpretation is supported by the plasma actuation spectrum in Figure~\ref{fig:controller_response2}(c), which distributes energy continuously across the disturbance band instead of concentrating it at isolated peaks.

Even under stochastic excitation, the cross-correlation in Figure~\ref{fig:controller_response2}(f) remains sharply peaked, showing that the learned control signal preserves a well-defined effective delay relative to the reference signal~($\bm{x_s}$). Compared with cases~$A$ and $B1$, the dominant lag extracted from the cross-correlation in case~$C1$ is smaller ($\Delta t \approx 183$ versus $\Delta t \approx 208$ and $214$, respectively). This indicates that, under broadband excitation, the learned FIR mapping is less well described as a delay-and-gain operator. Instead, the controller applies a frequency-dependent compromise across the unstable TS wave band, such that the cross-correlation peak should be interpreted as an effective delay of the broadband actuation strategy rather than as a direct estimate of a single convective time scale. The persistence of this delay structure across all control cases indicates that the SDRL framework consistently captures the convective character of the disturbance dynamics. At the same time, the comparison among cases~$A$, $B1$, and $C1$ shows a clear progression in the learned control strategy: from single-tone delay-and-gain compensation in case~$A$, to structured multi-tone compensation in case~$B1$, and finally to genuinely broadband actuation in case~$C1$.

\subsection{Controller performance}\label{Sec: controller_performance}

\begin{figure}
    \centering
    \includegraphics{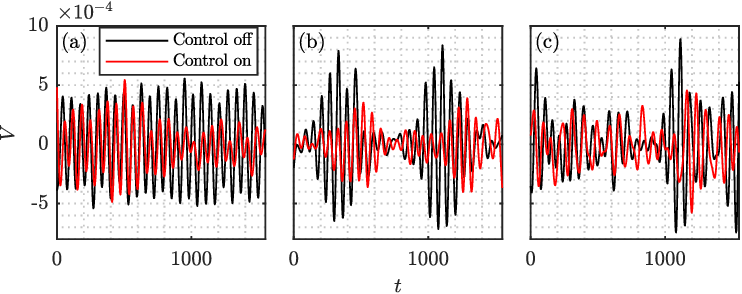}
    \captionsetup{justification=centering}
    \caption{Error signal~($\bm{e_s}$) time series corresponding to the best-performing FIR filter obtained for (a)~$A$, (b)~$B1$, and (c)~$C1$ control cases.}
   \label{fig:controller_performance1}
\end{figure}

\begin{figure}
    \centering
    \includegraphics{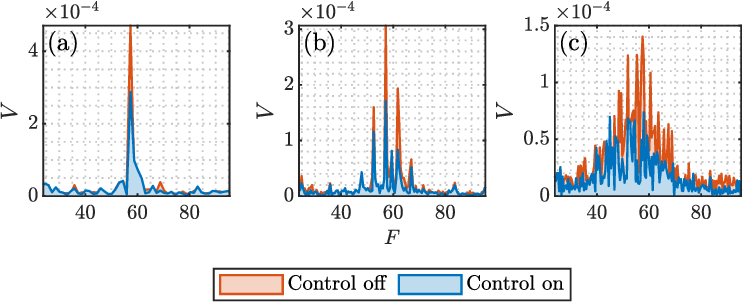}
    \captionsetup{justification=centering}
    \caption{Error signal~($\bm{e_s}$) Fourier transform corresponding to the best-performing FIR filter obtained for~$(a)~A$ , $(b)~B1$, and $(c)~C1$ control cases.}
   \label{fig:controller_performance2}
\end{figure}

The performance of the SDRL-based controller is evaluated by comparing the error signal~($\bm{e_s}$), evolution with and without control in the time and frequency domains, as shown in Figures~\ref{fig:controller_performance1} and~\ref{fig:controller_performance2}. The results correspond to the best-performing FIR filters obtained for ~$A$~(single-frequency), $B1$~(multi-frequency), and $C1$~(white-noise) control scenarios. Time is shown relative to the start of the plotted window ($t=0$ denotes the beginning of the extracted segment). The `control on' trace is taken from the final generations after convergence of the training, whereas the `control off' trace is a baseline acquired with the actuator off; since these were not recorded simultaneously, they do not correspond to the same realization of the disturbance and are compared in terms of amplitude statistics rather than pointwise phase.

Figure~\ref{fig:controller_performance1} displays the time series of the wall-pressure fluctuations measured at the error microphone. In all cases, the signal amplitude decreases after the controller is activated compared with the control-off condition, indicating effective attenuation of the incoming disturbances. For case~$A$ (Figure~\ref{fig:controller_performance1}(a)), the oscillations become significantly weaker under control, with a nearly sinusoidal waveform maintained at the excitation frequency of $57$. The reduction in amplitude confirms that the controller precisely counteracts the convected TS wave packet through the phase-aligned plasma actuation.

In case~$B1$~(Figure~\ref{fig:controller_performance1}(b)), the controlled signal exhibits a notable decrease in overall fluctuation level while retaining a more complex temporal structure due to the interaction of several frequency components. This indicates that the SDRL agent successfully identified a multi-tone control policy capable of attenuating multiple unstable modes simultaneously. The response remains stable and free from secondary amplification or oscillatory drift, confirming that the learned policy remains effective within the multi-frequency environment.

For case~$C1$ (Figure~\ref{fig:controller_performance1}(c)), the controller continues to suppress the pressure fluctuations despite the absence of a dominant frequency. The reduced amplitude envelope and the smoother temporal profile reflect the ability of the SDRL framework to extract relevant features from stochastic input and maintain robust opposition-type control even under random disturbance forcing.

The corresponding spectral content is presented in Figure~\ref{fig:controller_performance2}. The Fourier transforms of the error signals~($\bm{e_s}$) reveal substantial attenuation across the dominant frequency bands in all three scenarios.

\FloatBarrier

\FloatBarrier

In case~$A$~(Figure~\ref{fig:controller_performance2}(a)), the main spectral peak at $57$ decreases when the control is active. For case~$B1$~ (Figure~\ref{fig:controller_performance2}(b)), the controller achieves simultaneous reduction of multiple peaks between $52$ and $76$, maintaining consistent attenuation levels across the excited modes. In case~$C1$~(Figure~\ref{fig:controller_performance2}(c)), the broadband energy level of the controlled signal is reduced across frequencies, illustrating that the learned FIR filter behaves as an adaptive wideband compensator.

To provide a quantitative measure of the reductions observed in the spectra, the relative attenuation at the error sensor location~($x=58$) was evaluated and is summarized in Table~\ref{Table:attenuation_mic9}. It represents the normalized difference in the amplitude of each TS wave mode between the controlled and uncontrolled cases, with the normalization performed relative to the uncontrolled amplitude. In the broadband white-noise control case, the relative attenuation is computed as the mean percentage reduction of the Fourier amplitude spectrum between the controlled and uncontrolled signals,
\begin{equation}
A_{\mathrm{rel}} =
\left\langle
\frac{C_{\mathrm{off}}(F) - C_{\mathrm{on}}(F)}{C_{\mathrm{off}}(F)}
\right\rangle_{f \in \Omega}
\times 100.
\label{eq:white_noise_attenuation}
\end{equation}
where \(C_{\text{off}}\) and \(C_{\text{on}}\) are the spectral amplitudes with the controller off and on, respectively, and \(\Omega\) includes frequencies where \(C_{\text{off}}(F)\) exceeds the background noise threshold \(10^{-5}\,\mathrm{V}\). This provides a broadband measure of disturbance-energy reduction achieved by the controller.

In addition to peak-based attenuation measures, the control performance was quantified using a band-limited spectral energy-content metric, defined as the area under the squared amplitude spectrum within the disturbance frequency band of interest,
\begin{equation}
E = \int_{F_1}^{F_2} V(F)^2 \, \mathrm{d}F ,
\end{equation}
where $(F_1, F_2)=(36, 83)$ is the frequency
band where the laminar boundary layer exhibits pronounced TS wave amplification, Figure~\ref{fig:stability_curve}. The relative reduction was then computed as
\begin{equation}
\eta_E \equiv \frac{E_{\mathrm{off}} - E_{\mathrm{on}}}{E_{\mathrm{off}}} .
\end{equation}

\FloatBarrier

\begin{table}
\centering
\caption{Relative attenuation (\%) at the error microphone ($x=58$) for all control scenarios.}
\label{Table:attenuation_mic9}
\renewcommand{\arraystretch}{1.1}
\setlength{\tabcolsep}{7pt}
\begin{tabular}{ccccc}
\toprule
\toprule
\text{Control case} & \text{${Re_{\delta^{*},\mathrm{ref}}~(U_{\infty}[m/s])}$} & \text{F} & \text{$A_\mathrm{rel}$ [\%]} & \text{$\eta_E$ [\%]} \\
\midrule
$A$ & 1700~(20.0) & 57 & 40.5 & 53.89 \\
\midrule
\multirow{2}{*}{$B1$} & \multirow{2}{*}{1700~(20.0)} & 57 & 46.9 & \multirow{2}{*}{63.69} \\
                      &                                & 62 & 50.1 &       \\
\midrule
\multirow{3}{*}{$B2$} & \multirow{3}{*}{1912~(22.5)} & 57 & 61.9 & \multirow{3}{*}{76.92} \\
                      &                                & 62 & 45.6 &       \\
                      &                                & 67 & 62.2 &       \\
\midrule                                 
$C1$ & 1700~(20.0) & $36\text{--}83$ & 37.37 & 64.62 \\
\midrule                                 
$C2$ & 1912~(22.5) & $36\text{--}83$ & 39.32 & 68.07 \\
\bottomrule
\end{tabular}
\end{table}

Based on Table~\ref{Table:attenuation_mic9}, the controller consistently reduced disturbance amplitudes by $40$–$60\%$ at the error microphone location, performing best in case~$B2$~(multi-frequency excitation with higher freestream velocity). This trend is also reflected by the band-limited spectral energy reduction metric~($\eta_E$), which yields $53.89-76.92\%$. Among all cases, $B2$ therefore provides the strongest attenuation both in terms of discrete amplitude reduction and integrated spectral energy reduction over the selected band. Overall, the peak-based attenuation values suggest more effective suppression for narrowband disturbances, since the controller can synchronize phase and amplitude with a limited number of coherent modes rather than distributing control effort across a continuous spectrum. However, the band-limited spectral energy metric shows that the broadband cases still experience substantial integrated disturbance reduction, indicating that peak-based and energy-based measures emphasize different aspects of control performance.

The improved response at higher $U_{\infty}$ could arise from enhanced signal coherence and a more stable convection delay between sensors, which allows the controller to maintain accurate phase opposition. This tendency is consistent across both excitation classes considered in paired form, with the higher-velocity cases outperforming their lower-velocity counterparts in terms of band-limited spectral energy reduction. Although the improvement is more pronounced for the multi-frequency forcing, both comparisons support the conclusion that the controller benefits from the higher tested freestream velocity. These results indicate that the SDRL policy remains effective across the tested spectral conditions and freestream velocities. A dedicated comparison of the two tested freestream velocities is provided in Appendix~\ref{App:freestream}. 

The slightly lower performance observed in the case~$A$ can be attributed to two main factors. 
First, the controller tends to overfit when exposed to a purely monochromatic disturbance. 
In this scenario, the FIR filter rapidly converges toward a narrow phase–amplitude relationship specific to the excitation frequency, resulting in a highly tuned but fragile solution that loses effectiveness under minor frequency drifts or amplitude fluctuations. This can be seen by comparing the reference and plasma signal spectrum in Figure~\ref{fig:controller_response2}(a). This interpretation is also consistent with the band-limited energy metric, for which case $A$ yields the lowest reduction among all tested scenarios, further indicating that a purely monochromatic forcing environment leads to a less robust control solution.
Second, the multi-frequency forcing provides a richer spectral environment that enhances the learning dynamics of the SDRL algorithm. 
By interacting with several closely spaced tones, the controller is driven to identify a more general filter that minimizes the overall disturbance energy across the relevant frequency band, rather than optimizing for a single harmonic. 
This spectral diversity promotes smoother convergence, improved robustness, and a physically meaningful control law that remains effective under broader excitation conditions. 

\subsubsection{Flow-field validation using PIV (selected cases)}
\label{sec:piv_validation}

To verify that the microphone-based attenuation corresponds to a genuine reduction of disturbance energy within the boundary layer, two-component PIV measurements in the ($x,y$) mid-plane were performed in the downstream region indicated in Figure~\ref{fig:flow_problem}. These measurements complement the pressure-based assessment of controller performance and allow visualization of how the TS wave velocity field is modified by actuation. For each configuration, an ensemble of approximately $2000$ statistically independent, non-time-resolved velocity fields was acquired, such that the oscillatory phase is sampled randomly as described in Section~\ref{Sec:PIV-methodology}. Fluctuations were quantified from ensemble standard deviation fields of the velocity components, $\sigma_u(x,y)=\mathrm{std}[u(x,y,t)]$ and $\sigma_v(x,y)=\mathrm{std}[v(x,y,t)]$, providing robust estimates of the local RMS levels of streamwise and wall-normal perturbations. Two condensed metrics are reported: (i) streamwise-averaged wall-normal profiles, $\overline{\sigma_u}(y)$ and $\overline{\sigma_v}(y)$, obtained by averaging the $\sigma$-maps over the streamwise extent of the field of view; and (ii) a robust near-wall proxy based on the $90^{\mathrm{th}}$ percentile over $y$, denoted $P_{90}\{\sigma_u\}$ and $P_{90}\{\sigma_v\}(x)$, used to track the streamwise evolution of the strongest near-wall fluctuations. Since the wall-normal component typically carries higher uncertainty than the streamwise component, interpretation emphasizes $\sigma_u$. These statistics are reported for two control cases~$B1$ and $C1$ in Figure~\ref{fig:profile_PIV_combined_only_u}, while the case~$A$~(single-frequency) is provided in Appendix~\ref{App:PIV-appendix} for the sake of brevity. 

Figure~\ref{fig:profile_PIV_combined_only_u}~($a$) shows that the uncontrolled TS waves in case~$B1$ reach a streamwise velocity amplitude of approximately $1.85~\%~U_\infty$. Since TS wave amplitudes in the linear growth regime are typically below~$1\%$ of the freestream velocity, this amplitude level is more consistent with nonlinear development than with a strictly linear TS wave state. The measured streamwise disturbance profile also exhibits the characteristic dual-lobe structure of TS waves, with local maxima located at approximately $y = 1.0$ ($y/\delta_{99} = 0.33$) and $y = 3.0$ ($y/\delta_{99} = 1.0$), as reported also in some reference studies~\citep{grundmann2009experimental,widmann2012measuring}.

For case~$B1$, Figure~\ref{fig:profile_PIV_combined_only_u}~($a$--$b$) shows that actuation weakens the fluctuation levels near the wall and limits the downstream increase of $P_{90}\{\sigma_u/U_\infty\}$. Quantitatively, the peak value decreases from $2.15\,\%$ in the uncontrolled condition to $0.67\,\%$ under control, corresponding to a reduction of approximately $69\,\%$ measured at the end of the PIV domain. In addition, the area under the $P_{90}\{\sigma_u/U_\infty\}(x)$ curve over the common streamwise range of the PIV domain is reduced by approximately $57.3\,\%$. These results are consistent with the microphone-based attenuation reported in Table~\ref{Table:attenuation_mic9}, indicating that the attenuation measured at the sensors is accompanied by a substantial reduction in the streamwise velocity-fluctuation level throughout the boundary layer.

\begin{figure}
\centering
\includegraphics{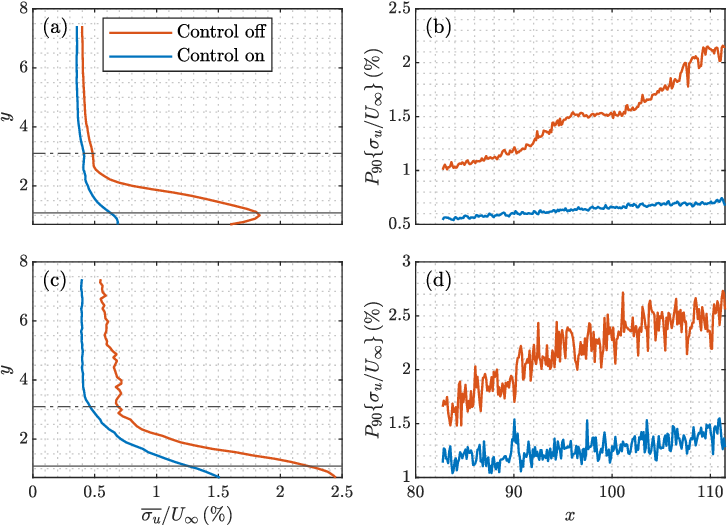}
\captionsetup{justification=centering}
\caption{Representative PIV fluctuation statistics: normalized wall-normal profiles of $\overline{\sigma_u}(y)$ and streamwise evolution of $P_{90}\{\sigma_u\}(x)$ for case~$B1$~($a-b$) and case~$C1$~($c-d$), comparing control off/on. Black lines
denote the displacement thickness~(solid) and the boundary-layer thickness~(dashed).}
\label{fig:profile_PIV_combined_only_u}
\end{figure}

Figure~\ref{fig:profile_PIV_combined_only_u}~($c$) shows that the uncontrolled disturbance field in case~$C1$ reaches an amplitude of approximately $2.5\,\%\,U_\infty$, indicating a more strongly nonlinear state than in case~$B1$. The streamwise disturbance profile nevertheless retains the characteristic two-lobe structure commonly associated with TS wave motion. However, the wall-normal positions of the two local maxima differ from those observed in case~$B1$. This difference is likely related to the broadband nature of the excitation in case~$C1$, for which the measured profile reflects the superposition of multiple amplified frequency components, together with stronger nonlinear development at the higher disturbance amplitude.

For case~$C1$, Figure~\ref{fig:profile_PIV_combined_only_u}~($c$--$d$) indicates that control also lowers the fluctuation levels over most of the measurement domain, with the clearest separation between the controlled and uncontrolled conditions appearing toward the downstream end of the PIV field of view. The peak $P_{90}\{\sigma_u/U_\infty\}$ decreases from $2.73\,\%$ without control to $1.26\,\%$ with control, i.e. by approximately $54\,\%$. Over the same streamwise interval, the area under the $P_{90}\{\sigma_u/U_\infty\}(x)$ profile is reduced by approximately $41.9\,\%$. Although this reduction is less uniform and less pronounced than in case~$B1$, the decrease in both the peak and the integrated fluctuation level shows that the learned FIR filter remains effective under broadband forcing.

\begin{figure}
\centering
\includegraphics{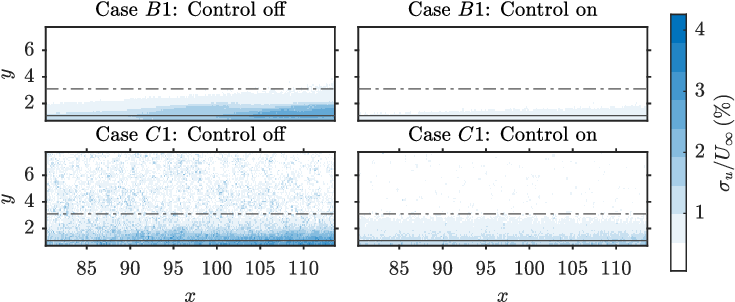}
\captionsetup{justification=centering}
\caption{Normalized streamwise velocity fluctuation field based on the standard deviation map ($\sigma_u$) for cases $B1$ and~$C1$. Black lines
denote the displacement thickness~(solid) and the boundary-layer thickness~(dashed).}
\label{fig:contour_PIV_combined}
\end{figure}

The spatial distribution of the fluctuations provides further insight into the control effect. Figure~\ref{fig:contour_PIV_combined} displays the normalized standard deviation fields of the streamwise velocity components for two selected control cases~$B1$ and ~$C1$. In all uncontrolled cases, $\sigma_u$ exhibits a distinct near-wall ridge that intensifies downstream, representing the convective growth of TS waves in the measurement region. With control active, this ridge weakens and becomes more diffuse, while the region of elevated $\sigma_u$ shrinks considerably. These spatial patterns confirm that the reduction observed at the microphones corresponds to a genuine decrease of fluctuation energy within the boundary layer.


Overall, the PIV diagnostics consistently show that the SDRL-based controller reduces the energy of the TS wave field downstream of the actuator. The effect is strongest for the streamwise velocity component and under multi-frequency forcing, moderate but sustained for broadband excitation, and concentrated near the wall where the instability attains its maximum amplitude. The spatial persistence of the reduced fluctuation levels confirms that the control waveform not only minimizes the local pressure fluctuations measured by the microphones but also weakens the instability as it convects downstream, contributing to a measurable delay of transition onset.

\subsection{Downstream persistence of disturbance attenuation and transition delay}

A key experimental finding is that the attenuation of disturbances persists downstream of the error sensor, demonstrating that the control effect extends beyond the measurement location. This persistence confirms that the SDRL-based controller not only minimizes pressure fluctuations locally but also weakens the TS waves as they convect further downstream, thereby delaying transition.

\begin{figure}
\centering
\includegraphics{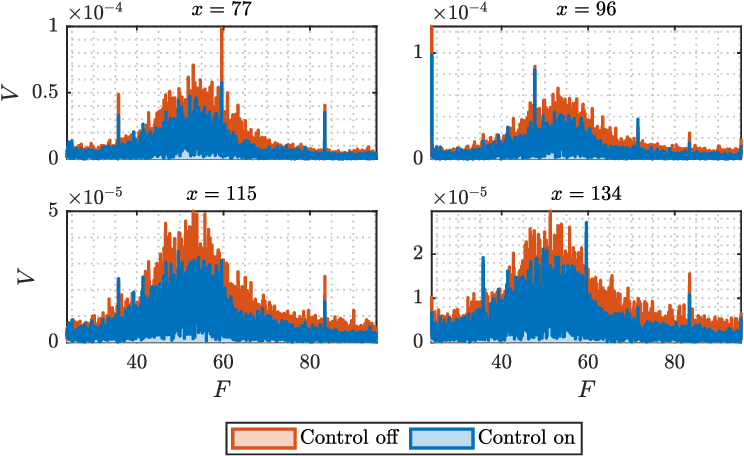}
\captionsetup{justification=centering}
\caption{Fourier transforms of the downstream microphone signals corresponding to the best-performing FIR filter obtained for~$C1$~(broadband white-noise) control case.}
\label{fig:error_downstream_WN}
\end{figure}

Figure~\ref{fig:error_downstream_WN} shows the Fourier spectra of the downstream microphones for the case~$C1$~(broadband white-noise). The spectra corresponding to the uncontrolled flow (orange) exhibit a broad distribution of amplified frequencies centered around $48-60$, consistent with the most unstable TS wave band predicted in the preliminary stability analysis in Figure~\ref{fig:stability_curve}. When the control is active (blue), the spectral energy in this range is markedly reduced at all downstream microphones, confirming that the reduction is maintained several wavelengths downstream of the actuator.

The persistence of attenuation downstream indicates that the control waveform generated by the optimized FIR filter interacts destructively with the convecting TS waves. The controller’s response effectively alters the local amplification rate of the disturbances, resulting in a net energy reduction within the boundary layer and, consequently, a downstream shift of the transition onset.

The consistent attenuation observed across multiple microphones provides clear experimental evidence that the algorithm actively manipulates the instability dynamics rather than merely suppressing the measurement signal. The sustained weakening of the disturbances is consistent with a downstream shift of the transition onset in these experiments.

To complement these qualitative observations, the effectiveness and spatial persistence of the control are quantified through the relative attenuation of the TS wave amplitude measured at all downstream microphones. The relative attenuation values presented in Table~\ref{Table: downstream_attenuation} follow the same definition used for Table~\ref{Table:attenuation_mic9}. This quantifies the reduction achieved by the controller at each microphone position, providing a direct percentage measure of control effectiveness.

\begin{table}
\centering
\caption{Relative attenuation (\%) at downstream microphone locations for all control scenarios. The relative attenuation for cases $C1$ and $C2$ is defined in Eq.~\eqref{eq:white_noise_attenuation}.}
\label{Table: downstream_attenuation}
\renewcommand{\arraystretch}{1.1}
\setlength{\tabcolsep}{5pt}
\begin{tabular}{cccccccc}
\toprule
\toprule
\text{Control case} & \text{$Re_{\delta^{*},\mathrm{ref}}$} & \text{F} &
$x=58$ & $x=77$ & $x=96$ & $x=115$ & $x=134$ \\
\midrule
$A$ & 1700 & 57 & 40.5 & 37.1 & 35.0 & 38.6 & 41.0 \\[2pt]
\midrule
\multirow{2}{*}{$B1$} & \multirow{2}{*}{1700}
 & 57 & 46.9 & 50.6 & 42.6 & 44.4 & 43.3 \\
 &  & 62 & 50.1 & 66.7 & 53.7 & 47.7 & 40.2 \\[2pt]
\midrule
\multirow{3}{*}{$B2$} & \multirow{3}{*}{1912}
 & 57 & 61.9 & 39.8 & 46.4 & 49.2 & 48.2 \\
 &  & 62 & 45.6 & 56.4 & 52.2 & 54.6 & 55.4 \\
 &  & 67 & 62.2 & 61.7 & 53.7 & 53.0 & 48.9 \\[2pt]

\midrule
$C1$ & 1700 & 36-83 & 37.37 & 37.59 & 41.46 & 41.23 & 47.42 \\
\midrule
$C2$ & 1912 & 36-83 & 39.32 & 42.25 & 38.41 & 41.91 & 51.05 \\

\end{tabular}
\end{table}

The attenuation levels summarized in Table~\ref{Table: downstream_attenuation} demonstrate that the SDRL-based controller achieved substantial TS wave reduction across all tested configurations. The downstream evolution of attenuation from the pressure signals ($x=58$) to $x=134$ provides insight into the persistence and spatial extent of the control effect.

At $Re_{\delta^{*},\mathrm{ref}} \approx 1700$~($U_{\infty} = 20~\text{m/s}$), case~$A$ exhibits a consistent reduction of approximately $35{-}41\%$ across all microphones, indicating that the controller effectively weakened the disturbance at the fundamental excitation frequency of $57$. The attenuation remains almost uniform along the streamwise direction, implying a stable and spatially coherent opposition between the actuation and the incoming TS wave. However, the magnitude of reduction is moderate compared with the multi-frequency and broadband cases, suggesting that the controller primarily compensated for a single dominant mode without addressing higher harmonics.

For~case~$B2$, attenuation levels reach their highest values, exceeding $60\%$ at the excitation frequencies. The strongest suppression occurs near $F=67$, where all microphones register reductions above $48\%$. The gradual spatial decay of attenuation downstream confirms that the learned filter produced a control signal capable of simultaneously canceling multiple components while maintaining phase coherence over the actuation region. 

When only two frequency components ($F=57$ and $62$) are present at the lower velocity, the overall attenuation remains in the range of $45{-}55\%$. This reduction is in accordance with the explanations in Section~\ref{Sec: controller_performance}. Under broadband excitation, attenuation levels remain between $38\%$ and $51\%$, comparable to the average of the deterministic cases. 

The nearly uniform response across frequencies and microphones suggests that the controller learned a generalized actuation pattern effective against a wide spectral content of disturbances. Although the attenuation is less localized around a single frequency peak, the broadband case confirms that the learned filter possesses spectral flexibility and maintains disturbance mitigation downstream.

Across most scenarios, attenuation is strongest at the microphones located at $x=58$ and $x=77$ and gradually decreases further downstream, indicating that disturbance suppression is most effective in the near-actuator region. The consistent positive attenuation across all microphones confirms that the control signal did not introduce amplification in any case. In summary, the SDRL controller achieved up to 62\% reduction of TS wave amplitude under multi-frequency forcing, maintained 45-50\% attenuation under reduced flow speed, and preserved robust performance under broadband excitation, demonstrating both adaptability and physical consistency of the learned opposition-control mechanism.

\section{Conclusions}\label{Sec: conclusion}
This study presented the first experimental implementation of a model-free single-step deep reinforcement learning (SDRL)-based controller for the suppression of Tollmien–Schlichting (TS) waves in a flat-plate boundary layer using dielectric-barrier discharge (DBD) plasma actuation. 
The controller autonomously learned, in real time, the finite-impulse-response (FIR) filter coefficients that map upstream wall-pressure fluctuations to the actuation command, thereby achieving adaptive opposition control without any explicit system identification or model reduction.

Across all disturbance scenarios, including single-frequency, multi-frequency, and broadband forcing, the SDRL controller consistently reduced the downstream disturbance energy. 
Attenuation levels reached up to \(62.2\%\) in TS wave amplitude under multi-frequency excitation, while broadband forcing yielded \(37\text{--}39\%\) amplitude reduction with corresponding spectral energy reductions of approximately \(65\text{--}68\%\), maintained under moderate variations in freestream velocity. 
The controller demonstrated rapid convergence, with policy stabilization within a few minutes of operation, and the learned filter exhibited physically interpretable amplitude–phase characteristics aligned with the convective delay between sensors. 
Downstream measurements confirmed that the control effect persisted several wavelengths beyond the actuator, signifying a genuine reduction of disturbance energy and supporting transition-delay potential.

The results establish that the SDRL framework can deliver robust, interpretable, and real-time flow-control performance within the practical constraints of wind-tunnel experimentation. 
Its compact formulation, low computational cost, and strong adaptability make it a promising alternative to conventional adaptive schemes, particularly for applications where explicit modeling or system identification is impractical.

Future research could focus on extending the present framework toward a multi-step deep reinforcement learning algorithm to enhance the controller’s capability in terms of (i) learning state-dependent and history-aware policies through proper temporal credit assignment; (ii) coordinating multiple sensors and actuators, including spanwise-varying actuation; and (iii) establishing robustness guarantees under variations in $U_{\infty}$, Reynolds number, and adverse pressure-gradient conditions.

\bibliographystyle{jfm}

\bibliography{merged_unique_references}

@PREAMBLE{
 "\providecommand{\noopsort}[1]{}" 
 # "\providecommand{\singleletter}[1]{#1}%" 
}

@article{Walther2001OptimalLayer,
    title = {{Optimal control of tollmien-schlichting waves in a developing boundary layer}},
    year = {2001},
    journal = {Physics of Fluids},
    author = {Walther, S. and Airiau, C. and Bottaro, A.},
    number = {7},
    pages = {2087--2096},
    volume = {13},
    publisher = {American Institute of Physics Inc.},
    doi = {10.1063/1.1378035},
    issn = {10706631}
}

@book{Mayes2003Boundary-LayerTheory,
    title = {{Boundary-Layer Theory}},
    year = {2003},
    booktitle = {Physic and astronomy},
    author = {Mayes, C and Schlichting, H and Krause, E and Oertel, H J and Gersten, K},
    publisher = {Springer}
}

@article{Thomas,
  author    = {Andrew S. W. Thomas},
  title     = {The Control of Boundary-Layer Transition Using a Wave-Superposition Principle},
  journal   = {Journal of Fluid Mechanics},
  pages     = {233--250},
  year      = {1983},
}

@article{Marios2015,
   author = {M. Kotsonis and Ram Krishan Shukla and Stefan Pröbsting},
   issn = {17568250},
   issue = {1-2},
   journal = {International Journal of Flow Control},
   month = {6},
   pages = {37-54},
   publisher = {Multi-Science Publishing Co. Ltd},
   title = {Control of natural {Tollmien-Schlichting} waves using dielectric barrier discharge plasma actuators},
   volume = {7},
   year = {2015},
}

@article{rabault2019artificial,
  title={Artificial neural networks trained through deep reinforcement learning discover control strategies for active flow control},
  author={Rabault, J. and Kuchta, M. and Jensen, A. and Réglade, U. and Ceradi, N.},
  journal={Journal of Fluid Mechanics},
  volume={865},
  pages={281--302},
  year={2019}
}

@article{rabault2019accelerating,
  title={Accelerating deep reinforcement learning strategies of flow control through a multi-environment approach},
  author={Rabault, Jean and Kuhnle, Alexander},
  journal={Physics of Fluids},
  volume={31},
  pages={094105},
  year={2019}
}

@Article{Gad3,
  author = 	 {Gad-El-Hak, M},
  title = 	 {MODERN DEVELOPMENTS IN FLOW CONTROL},
  journal = 	 {Applied Mechanics Reviews},
  volume = 	 {49},
  issue = {2},
  year = {1996},
}

@article{superposition3, title={Active control of laminar-turbulent transition}, volume={118}, journal={Journal of Fluid Mechanics}, publisher={Cambridge University Press}, author={Liepmann, H. W. and Nosenchuck, D. M.}, year={1982}, pages={201–204}}

@article{modelfree2,
  title={Active cancellation of {Tollmien–Schlichting} instabilities on a wing using multi-channel sensor actuator systems},
  author={Dana J. Sturzebecher and Wolfgang Hartmut Nitsche},
  journal={International Journal of Heat and Fluid Flow},
  year={2003},
  volume={24},
  pages={572-583},

}

@article{modelfree3,
   author = {A. Kurz and N. Goldin and R. King and C. Tropea and S. Grundmann},
   issn = {07234864},
   issue = {11},
   journal = {Experiments in Fluids},
   month = {12},
   title = {Hybrid transition control approach for plasma actuators},
   volume = {54},
   year = {2013},
}

@article{Ladd1988,
  author = {Ladd, D. and Hendricks, E.},
  year = {1988},
  title = {Active control of {2-D} instability waves on an axisymmetric body},
  journal = {Exp. Fluids},
  volume = {6},
  pages = {69--70}
}

@inproceedings{baumann1996,
  author = {Baumann, M. and Nitsche, W.},
  title = {Investigation of Active Control of {Tollmien-Schlichting} Waves on a Wing},
  booktitle = {Transitional Boundary Layers in Aeronautics},
  volume = {46},
  editor = {Henkes, R. and van Ingen, J.},
  publisher = {KNAW, Amsterdam, Netherlands},
  pages = {89--98},
  year = {1996},
}

@article{albrecht2006,
  author = {Albrecht, T. and Grundmann, R. and Mutschke, G. and Gerbeth, G.},
  title = {On the Stability of the Boundary Layer Subject to a Wall-Parallel {Lorentz} Force},
  journal = {Physics of Fluids},
  volume = {18},
  number = {9},
  pages = {098103},
  year = {2006},
}

@article{belson2013,
  author = {Belson, B. A. and Semeraro, O. and Rowley, C. W. and Henningson, D. S.},
  title = {Feedback control of instabilities in the two-dimensional {{Blasius}} boundary layer: The role of sensors and actuators},
  journal = {Physics of Fluids},
  volume = {25},
  number = {5},
  year = {2013},
}

@article{chen2023,
  author = {Chen, W. and Wang, Q. and Yan, L. and Hu, G. and Noack, B. R.},
  title = {Deep reinforcement learning-based active flow control of vortex-induced vibration of a square cylinder},
  journal = {Physics of Fluids},
  volume = {35},
  number = {5},
  year = {2023},
}

@article{guastoni2023,
  author = {Guastoni, L. and Rabault, J. and Schlatter, P. and Azizpour, H. and Vinuesa, R.},
  title = {Deep reinforcement learning for turbulent drag reduction in channel flows},
  journal = {European Physical Journal E},
  volume = {46},
  number = {4},
  year = {2023},
}

@article{li2022,
  author = {Li, J. and Zhang, M.},
  title = {Reinforcement-learning-based control of confined cylinder wakes with stability analyses},
  journal = {Journal of Fluid Mechanics},
  volume = {932},
  year = {2022},
}

@article{ren2021,
  author = {Ren, F. and Rabault, J. and Tang, H.},
  title = {Applying deep reinforcement learning to active flow control in weakly turbulent conditions},
  journal = {Physics of Fluids},
  volume = {33},
  number = {3},
  year = {2021},
}

@article{Viquerat2022,
   author = {J. Viquerat and R. Duvigneau and P. Meliga and A. Kuhnle and E. Hachem},
   issn = {14333058},
   journal = {Neural Computing and Applications},
   month = {1},
   publisher = {Springer Science and Business Media Deutschland GmbH},
   title = {Policy-based optimization: single-step policy gradient method seen as an evolution strategy},
   volume = {35},
   year = {2022},
}

@article{Ghraieb2021,
   author = {H. Ghraieb and J. Viquerat and A. Larcher and P. Meliga and E. Hachem},
   issn = {2469990X},
   issue = {5},
   journal = {Physical Review Fluids},
   month = {5},
   publisher = {American Physical Society},
   title = {Single-step deep reinforcement learning for open-loop control of laminar and turbulent flows},
   volume = {6},
   year = {2021},
}

@article{Ghraieb2022,
   author = {H. Ghraieb and J. Viquerat and A. Larcher and P. Meliga and E. Hachem},
   issn = {21583226},
   issue = {8},
   journal = {AIP Advances},
   month = {8},
   publisher = {American Institute of Physics Inc.},
   title = {Single-step deep reinforcement learning for two- and three-dimensional optimal shape design},
   volume = {12},
   year = {2022},
}

@article{Viquerat2021,
   author = {Jonathan Viquerat and Jean Rabault and Alexander Kuhnle and Hassan Ghraieb and Aurélien Larcher and Elie Hachem},
   issn = {10902716},
   journal = {Journal of Computational Physics},
   month = {3},
   publisher = {Academic Press Inc.},
   title = {Direct shape optimization through deep reinforcement learning},
   volume = {428},
   year = {2021},
}

@inproceedings{mohammadikalakoo_canada,
  author = {Mohammadikalakoo, B. and Kotsonis, M. and Doan, N. A. K.},
  title = {{Optimization of {Tollmien-Schlichting} waves control: comparison between deep reinforcement learning and particle swarm optimization approach}},
  booktitle = {Thirteenth International Symposium on Turbulence and Shear Flow Phenomena (TSFP13)},
  address = {Montréal, Canada},
  year = {2024},
}

@article{mohammadikalakoo_paper,
    author = {Mohammadikalakoo, B. and Kotsonis, M. and Doan, N. A. K.},
    title = {Control of {Tollmien–Schlichting} waves using particle swarm optimization},
    journal = {Physics of Fluids},
    volume = {36},
    number = {12},
    year = {2024},
    month = {12},
    issn = {1070-6631},
}

@book{Schlichting2017,
  author    = {Hermann Schlichting and Klaus Gersten},
  title     = {Boundary-Layer Theory},
  edition   = {9},
  publisher = {Springer},
  address   = {Berlin/Heidelberg},
  year      = {2017}
}

@article{Paris2021,
  author    = {R. Paris and S. Beneddine and J. Dandois},
  title     = {Robust flow control and optimal sensor placement using deep reinforcement learning},
  journal   = {Journal of Fluid Mechanics},
  volume    = {913},
  pages     = {A25},
  year      = {2021},
}

@article{Font2025,
   author = {Bernat Font and Francisco Alcántara-Ávila and Jean Rabault and Ricardo Vinuesa and Oriol Lehmkuhl},
   issn = {2041-1723},
   issue = {1},
   journal = {Nature Communications},
   month = {2},
   pages = {1422},
   title = {Deep reinforcement learning for active flow control in a turbulent separation bubble},
   volume = {16},
   year = {2025},
}

@article{Hu2024,
   author = {Wulong Hu and Zhangze Jiang and Mingyang Xu and Hanyu Hu},
   issn = {10897666},
   issue = {7},
   journal = {Physics of Fluids},
   month = {7},
   publisher = {American Institute of Physics},
   title = {Efficient deep reinforcement learning strategies for active flow control based on physics-informed neural networks},
   volume = {36},
   year = {2024},
}

@article{Brito2021,
  author    = {Brito, P. P. C. and Morra, P. and Cavalieri, A. V. G. and Jordan, P. and Luchtenburg, D. M. and Sipp, D. and Agarwal, A. and Oliveira, A. P. L.},
  title     = {Experimental control of {Tollmien–Schlichting} waves using pressure sensors and plasma actuators},
  journal   = {Experiments in Fluids},
  volume    = {62},
  pages     = {32},
  year      = {2021},
}

@article{dorr2018,
author = {D\"{o}rr, Philipp C. and Kloker, Markus J.},
title = {Numerical Investigations on {Tollmien–Schlichting} Wave Attenuation Using Plasma-Actuator Vortex Generators},
journal = {AIAA Journal},
volume = {56},
number = {4},
pages = {1305-1309},
year = {2018},
}

@techreport{Mack84,
  author = {L. M. Mack},
  title = {Boundary-layer linear stability theory},
  number = {709},
  institution = {AGARD},
  year = {1984},
  type         = {Technical Report},
  institution  = {Jet Propulsion Laboratory, California Institute of Technology},
}

@article{linot2023,
  title={Turbulence control in plane Couette flow using low-dimensional neural ODE-based models and deep reinforcement learning},
  author={Linot, Alexander J. and Zeng, Kaidi and Graham, Michael D.},
  journal={International Journal of Heat and Fluid Flow},
  volume={101},
  pages={109136},
  year={2023},
  publisher={Elsevier},
  doi={10.1016/j.ijheatfluidflow.2022.109136}
}

@article{Babak2025,
  title={Real-time single-step deep reinforcement learning framework for control of Tollmien-Schlichting waves},
  author={Mohammadikalakoo, B and Kotsonis, M and Doan, NAK},
  journal={Physical Review Fluids},
  volume={10},
  number={12},
  pages={124902},
  year={2025},
  publisher={APS}
}

@article{MerinoMartinez2020,
  author    = {R. Merino-Mart{\'i}nez and A. Rubio Carpio and L. T{\'e}rcio and L. Pereira 
               and S. van Herk and F. Avallone and M. Ragni and M. Kotsonis},
  title     = {Aeroacoustic design and characterization of the 3D-printed, open-jet, anechoic wind tunnel of Delft University of Technology},
  journal   = {Applied Acoustics},
  volume    = {170},
  pages     = {107504},
  year      = {2020},
  doi       = {10.1016/j.apacoust.2020.107504}
}

@incollection{Lin1992,
  author    = {N. Lin and H. L. Reed and W. S. Saric},
  title     = {Effect of leading-edge geometry on boundary-layer receptivity to freestream sound},
  booktitle = {Instability, Transition, and Turbulence},
  publisher = {Springer},
  year      = {1992},
  pages     = {421--440}
}

@article{schultz2007rough,
  title={The rough-wall turbulent boundary layer from the hydraulically smooth to the fully rough regime},
  author={Schultz, MP and Flack, KA},
  journal={Journal of fluid mechanics},
  volume={580},
  pages={381--405},
  year={2007},
  publisher={Cambridge University Press}
}

@inproceedings{saric1983effect,
  author       = {Saric, W. S. and Reed, H. L.},
  title        = {Effect of Suction and Blowing on {Boundary-Layer} Transition},
  booktitle    = {Proceedings of the 21st Aerospace Sciences Meeting},
  address      = {Reno, NV, USA},
  month        = jan,
  year         = {1983},
  organization = {American Institute of Aeronautics and Astronautics (AIAA)},
  note         = {AIAA Paper 83-0043},
  doi          = {10.2514/6.1983-43}
}

@article{tol2019experimental,
  title={Experimental model-based estimation and control of natural Tollmien--Schlichting waves},
  author={Tol, HJ and De Visser, CC and Kotsonis, Marios},
  journal={AIAA Journal},
  volume={57},
  number={6},
  pages={2344--2355},
  year={2019},
  publisher={American Institute of Aeronautics and Astronautics}
}

@article{grundmann2009experimental,
  title={Experimental damping of boundary-layer oscillations using DBD plasma actuators},
  author={Grundmann, Sven and Tropea, Cameron},
  journal={International Journal of Heat and Fluid Flow},
  volume={30},
  number={3},
  pages={394--402},
  year={2009},
  publisher={Elsevier}
}

@article{widmann2012measuring,
  title={Measuring Tollmien--Schlichting waves using phase-averaged particle image velocimetry},
  author={Widmann, Alexander and Duchmann, Alexander and Kurz, Armin and Grundmann, Sven and Tropea, Cameron},
  journal={Experiments in Fluids},
  volume={53},
  number={3},
  pages={707--715},
  year={2012},
  publisher={Springer}
}

@article{wieneke2015piv,
  title={PIV uncertainty quantification from correlation statistics},
  author={Wieneke, Bernhard},
  journal={Measurement Science and Technology},
  volume={26},
  number={7},
  pages={074002},
  year={2015},
  publisher={IOP Publishing}
}

@article{sciacchitano2019uncertainty,
  title={Uncertainty quantification in particle image velocimetry},
  author={Sciacchitano, Andrea},
  journal={Measurement Science and Technology},
  volume={30},
  number={9},
  pages={092001},
  year={2019},
  publisher={IOP Publishing}
}

@article{carpenter85,
  title={The hydrodynamic stability of flow over Kramer-type compliant surfaces. Part 1. Tollmien-Schlichting instabilities},
  author={Carpenter, PW and Garrad, AD},
  journal={Journal of Fluid Mechanics},
  volume={155},
  pages={465--510},
  year={1985},
  publisher={Cambridge University Press}
}

@article{davies1997,
  author = {C. Davies and P. W. Carpenter},
  title = {Numerical simulation of the evolution of Tollmien–Schlichting waves over finite compliant panels},
  journal = {Journal of Fluid Mechanics},
  volume = {352},
  pages = {205--243},
  year = {1997},
  doi = {10.1017/S0022112097007606}
}

@article{hussein2015,
  title={Flow stabilization by subsurface phonons},
  author={Hussein, Mahmoud I and Biringen, Sedat and Bilal, Osama R and Kucala, Alec},
  journal={Proceedings of the Royal Society A: Mathematical, Physical and Engineering Sciences},
  volume={471},
  number={2177},
  pages={20140928},
  year={2015},
  publisher={The Royal Society Publishing}
}

@article{michelis2023,
  author = {T. Michelis and A. B. Putranto and M. Kotsonis},
  title = {Attenuation of Tollmien–Schlichting waves using resonating surface-embedded phononic crystals},
  journal = {Physics of Fluids},
  volume = {35},
  number = {4},
  pages = {044101},
  year = {2023},
  doi = {10.1063/5.0146795}
}

@article{michelis2023_2,
  title={On the interaction of Tollmien--Schlichting waves with a wall-embedded Helmholtz resonator},
  author={Michelis, Theodorus and De Koning, C and Kotsonis, M},
  journal={Physics of Fluids},
  volume={35},
  number={3},
  year={2023},
  publisher={AIP Publishing}
}

@article{Bagheri2009-input,
   title = {Input-output analysis, model reduction and control of the flat-plate boundary layer},
   author = {Shervin Bagheri and Luca Brandt and Dan S. Henningson},
   journal = {Journal of Fluid Mechanics},   
   doi = {10.1017/S0022112008004394},
   issn = {14697645},
   pages = {263-298},
   publisher = {Cambridge University Press},
   volume = {620},
   year = {2009},
}

@inproceedings{roth98,
  title={Boundary layer flow control with a one atmosphere uniform glow discharge surface plasma},
  author={Roth, J and Sherman, Daniel and Wilkinson, Stephen},
  booktitle={36th AIAA Aerospace Sciences Meeting and Exhibit},
  pages={328},
  year={1998}
}

@article{fabbiane2017,
  title={Energy efficiency and performance limitations of linear adaptive control for transition delay},
  author={Fabbiane, Nicol{\`o} and Bagheri, Shervin and Henningson, Dan S},
  journal={Journal of Fluid Mechanics},
  volume={810},
  pages={60--81},
  year={2017},
  publisher={Cambridge University Press}
}

@article{duchmann2013,
  title={Delay of natural transition with dielectric barrier discharges},
  author={Duchmann, A and Grundmann, S and Tropea, Cameron},
  journal={Experiments in fluids},
  volume={54},
  number={3},
  pages={1461},
  year={2013},
  publisher={Springer}
}

@article{Fransson2004,
  author       = {J. H. M. Fransson and L. Brandt and A. Talamelli and C. Cossu},
  title        = {Experimental study of the stabilizing effect of streaks on Tollmien–Schlichting waves},
  journal      = {Physics of Fluids},
  volume       = {16},
  number       = {10},
  pages        = {3627--3638},
  year         = {2004},
  doi          = {10.1063/1.1774152}
}

@article{fransson2006,
  title={Delaying transition to turbulence by a passive mechanism},
  author={Fransson, Jens HM and Talamelli, Alessandro and Brandt, Luca and Cossu, Carlo},
  journal={Physical review letters},
  volume={96},
  number={6},
  pages={064501},
  year={2006},
  publisher={APS}
}

@inproceedings{fan1993active,
  title={Active flow control with neural networks},
  author={Fan, Xuetong and Hofmann, Lorenz and Herbert, Thorwald},
  booktitle={3rd Shear Flow Conference},
  pages={3273},
  year={1993}
}

@inproceedings{fan1995transition,
  title={Transition control with neural networks},
  author={Fan, Xuetong and Herbert, Thorwald and Haritonidis, Joseph},
  booktitle={33rd Aerospace Sciences Meeting and Exhibit},
  pages={674},
  year={1995}
}

@article{wang2023,
  author = {Wang, C. and Yu, P. and Huang, H.},
  title = {Reinforcement-learning-based parameter optimization of a splitter plate downstream in cylinder wake with stability analyses},
  journal = {Physical Review Fluids},
  volume = {8},
  number = {8},
  year = {2023},

  
}

@article{fabbiane2014,
  title={Adaptive and model-based control theory applied to convectively unstable flows},
  author={Fabbiane, Nicolo and Semeraro, Onofrio and Bagheri, Shervin and Henningson, Dan S},
  journal={Applied Mechanics Reviews},
  volume={66},
  number={6},
  pages={060801},
  year={2014},
  publisher={American Society of Mechanical Engineers}
}

@techreport{Liepmann1946,
  author      = {H. W. Liepmann and G. H. Fila},
  title       = {Investigations of Effects of Surface Temperature and Single Roughness Elements on Boundary-Layer Transition},
  institution = {NACA},
  number      = {Report No. 890},
  year        = {1946},
  pages       = {587--598}
}

@phdthesis{Nosenchuck1982,
  author       = {D. M. Nosenchuck},
  title        = {Passive and Active Control of Boundary Layer Transition},
  school       = {California Institute of Technology},
  year         = {1982},
  type         = {Dissertation}
}
\appendix
\section{Perturbation velocity measurements}\label{App:PIV-appendix}

This appendix complements the discussion in the main text by providing the perturbation velocity statistics extracted from the PIV measurements. The analysis is based on $2000$ instantaneous PIV snapshots acquired for each condition, from which the velocity fluctuations are quantified using the standard deviations of the streamwise and wall-normal components, denoted by $\sigma_u$ and $\sigma_v$. To compactly represent the fluctuation level, two views are reported: (i) mean wall-normal profiles, $\overline{\sigma_u}(y)$ and $\overline{\sigma_v}(y)$, and (ii) a streamwise evolution measure based on the $90^{\mathrm{th}}$ percentile, $P_{90}\{\sigma_u\}$ and $P_{90}\{\sigma_v\}$, evaluated along~$x$. In addition, spatial distributions of $\sigma_u$ and $\sigma_v$ are shown to illustrate how the disturbance field evolves within the measurement plane.

\begin{figure}
\centering
\includegraphics{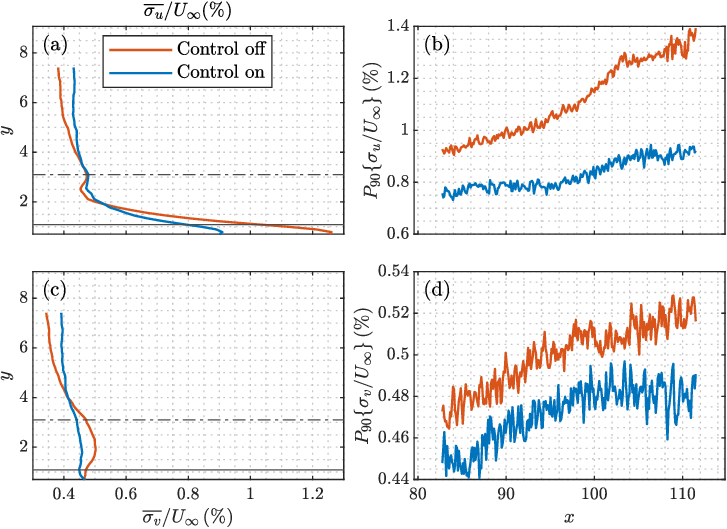}
\captionsetup{justification=centering}
\caption{Perturbation velocity statistics for case~$A$.~$(a,c)$: normalized mean wall-normal profiles $\overline{\sigma_u}(y)$ and $\overline{\sigma_v}(y)$.~$(b,d)$: normalized streamwise evolution of the $90^{\mathrm{th}}$ percentile, $P_{90}\{\sigma_u\}$ and $P_{90}\{\sigma_v\}$. Black lines
denote the displacement thickness~(solid) and the boundary-layer thickness~(dashed).}
\label{fig:profile_PIV_single}
\end{figure}

Figure~\ref{fig:profile_PIV_single} summarizes the perturbation velocity statistics for the case~$A$. Activating the controller leads to a clear reduction in the fluctuation level throughout the boundary layer. In the wall-normal direction, the mean profiles $\overline{\sigma_u}(y)$ and $\overline{\sigma_v}(y)$ decrease across the measured region, with the most pronounced attenuation occurring close to the wall, where the TS wave amplitude typically reaches its maximum. Consistently, the streamwise evolution of $P_{90}\{\sigma_u\}$ becomes noticeably flatter than in the uncontrolled flow, indicating that the SDRL-based controller mitigates the downstream amplification of TS disturbances. The reduction in $P_{90}\{\sigma_v\}$ is more modest, which is consistent with the predominantly streamwise nature of TS fluctuations and with the higher relative uncertainty associated with the wall-normal velocity component.

\begin{figure}
\centering
\includegraphics{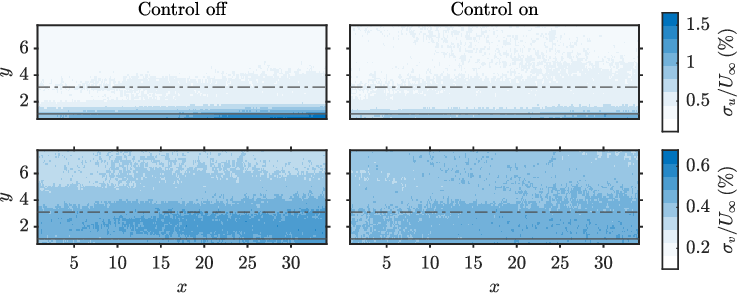}
\captionsetup{justification=centering}
\caption{Normalized standard deviation fields of the perturbation velocities, $\sigma_u$ and $\sigma_v$, for case~$A$. Black lines
denote the displacement thickness~(solid) and the boundary-layer thickness~(dashed).}
\label{fig:contour_PIV_single}
\end{figure}

Figure~\ref{fig:contour_PIV_single} shows the corresponding standard-deviation fields. The $\sigma_u$ map exhibits the strongest attenuation in the near-wall region and along the disturbance path, whereas the $\sigma_v$ map follows the same qualitative trends but with smaller magnitudes.

\section{Effect of Freestream Velocity} \label{App:freestream}

This appendix briefly examines the effect of freestream velocity on the disturbance development and the controller response. As $U_{\infty}$ increases, the boundary layer becomes thinner, and the amplification rate of the TS waves rises due to the higher local Reynolds number. The most amplified frequency band shifts slightly toward higher values, consistent with the scaling of the phase velocity with $U_{\infty}$. Moreover, the convective time delay between sensors decreases as the disturbances are advected faster downstream. These changes in the base-flow stability characteristics make the higher-speed condition somewhat more unstable and therefore potentially more demanding for control.

To preliminarily assess the sensitivity of the SDRL-based controller to such variations, two freestream velocities were tested: $Re_{\delta^{*},\mathrm{ref}} \approx 1700$~($U_{\infty} = 20~\text{m/s}$) and $Re_{\delta^{*},\mathrm{ref}} \approx 1912$~($U_{\infty} = 22.5~\text{m/s}$). The change in $U_{\infty}$ alters both the temporal correlation between sensors and the attainable control authority, thereby providing an opportunity to evaluate how the controller adapts to different convective timescales and instability growth rates.

\begin{figure}
    \centering
    \includegraphics{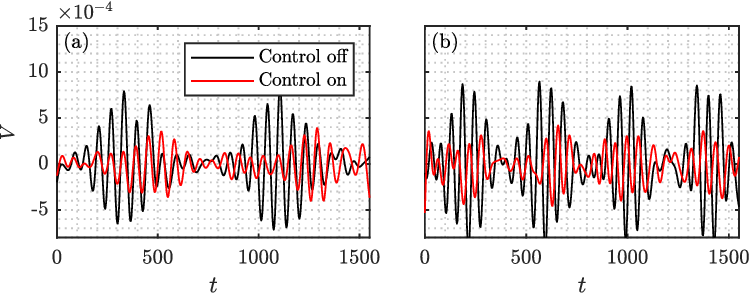}
    \captionsetup{justification=centering}
    \caption{Error signal~($\bm{e_s}$) time series corresponding to the best-performing FIR filter obtained for cases~(a)~$B1$~at~$Re_{\delta^{*},\mathrm{ref}} \approx 1700$~($U_{\infty} = 20~\text{m/s}$)~and~(b)~$B2$~at~$Re_{\delta^{*},\mathrm{ref}} \approx 1912$~($U_{\infty} = 22.5~\text{m/s}$).}
    \label{fig:freestream_effect2}
\end{figure}

\begin{figure}
    \centering
    \includegraphics{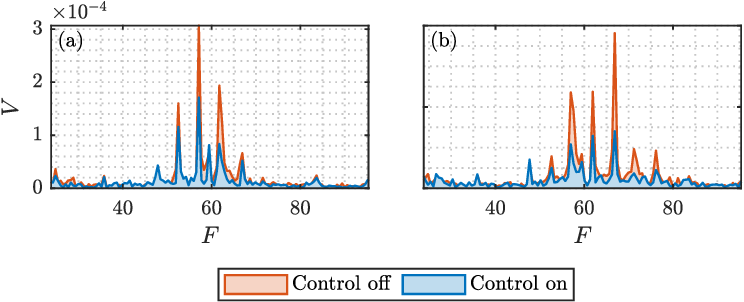}
    \captionsetup{justification=centering}
    \caption{Error signal~($\bm{e_s}$) Fourier transform corresponding to the best-performing FIR filter obtained for cases~(a)~$B1$~at~$Re_{\delta^{*},\mathrm{ref}} \approx 1700$~($U_{\infty} = 20~\text{m/s}$)~and~(b)~$B2$~at~$Re_{\delta^{*},\mathrm{ref}} \approx 1912$~($U_{\infty} = 22.5~\text{m/s}$).}
    \label{fig:freestream_effect}
\end{figure}

At $Re_{\delta^{*},\mathrm{ref}} \approx 1700$, the controller exhibited stable convergence and a clear attenuation of the disturbance energy within the instability band~$36-83$ as shown in Figures~\ref{fig:freestream_effect2} and \ref{fig:freestream_effect}. The optimized FIR filter achieved attenuation levels of approximately $40$–$55\%$ at the error microphone. When the freestream velocity increases, the boundary layer becomes thinner, and the amplification rate of the disturbances increases. Despite these changes, the SDRL-based controller maintains comparable attenuation performance, illustrated in Figures~\ref{fig:freestream_effect2} and \ref{fig:freestream_effect}.



Figure~\ref{fig:freestream_effect2} shows the time traces of the error signal~($\bm{e_s}$) for both velocities. The amplitude of the pressure fluctuations is notably reduced when control is active, with the oscillations at $U_{\infty} = 22.5~\text{m/s}$ following a similar decay trend as those observed at $20~\text{m/s}$. The controller continues to produce a stable and phase-aligned response despite the shorter convective timescale. The corresponding frequency-domain representation in Figure~\ref{fig:freestream_effect} confirms that the dominant TS wave peaks remain attenuated at both flow conditions, demonstrating that moderate variations in $U_{\infty}$ do not substantially degrade control effectiveness.

Similar behavior is observed for cases~$C1-C2$ in these two freestream velocities as shown in Tables~\ref{Table:attenuation_mic9} and~\ref{Table: downstream_attenuation}. It must be emphasized that these results are based on only two test velocities and, therefore cannot be interpreted as conclusive evidence of robustness. A systematic investigation over a broader range of $U_{\infty}$ would be required to quantify the sensitivity of the control law and to determine possible limits of its generalization capability. Nonetheless, the present findings suggest that moderate variations in $U_{\infty}$ do not drastically alter the control behavior, providing an encouraging first indication that the controller can adapt to small changes in convective timescale and boundary layer stability.

\end{document}